%% file: main.tex
\documentclass[fleqn,usenatbib]{mnras}

\usepackage{newtxtext,newtxmath}

\usepackage[T1]{fontenc}

\DeclareRobustCommand{\VAN}[3]{#2}
\let\VANthebibliography\thebibliography
\def\thebibliography{\DeclareRobustCommand{\VAN}[3]{##3}\VANthebibliography}

\usepackage{booktabs}
\usepackage{graphicx}	
\usepackage{amsmath}	
\usepackage{enumerate}
\usepackage{subcaption}
\usepackage{multicol}        
\usepackage{bm}		
\usepackage{pdflscape}	

\usepackage{xcolor}
\usepackage{ulem}

\usepackage[T1]{fontenc}
\usepackage{ae,aecompl}

\usepackage{newtxtext,newtxmath}

\usepackage{hyperref}
\usepackage{enumerate} 

\usepackage{microtype}

\input definitions

\definecolor{magenta}{RGB}{204, 00, 204}
\definecolor{darkgreen}{RGB}{00, 152, 26}
\definecolor{darkorange}{RGB}{205, 90, 0}

\newcommand{\umuG}{\,\mu\mathrm{G}}
\newcommand{\ukms}{\ensuremath{\,\mathrm{km\,s^{-1}}}}
\newcommand{\uicmc}{\ensuremath{\,\mathrm{cm}}^{-3}}

\title[Synthetic polarimetry of magnetized pillars]{Synthetic polarization observations of magnetized pillars in H\,{\sc ii} regions: Assessing the reliability of the Davis–Chandrasekhar–Fermi method}

\author[Hern{\'a}ndez-Cruz et al.]{Luis Andr{\'e}s Hern{\'a}ndez-Cruz,$^{1}$ 
Manuel Zamora-Avil{\'e}s,$^{2,1}$ 
Abraham Luna,$^{1}$ Ra{\'u}l Naranjo-Romero,$^{1}$ 
\and Jos{\'e} Franco,$^{3}$\thanks{Contact e-mail: \href{pepe@astro.unam.mx}{pepe@astro.unam.mx}}
Aina Palau,$^{4}$ Alejandro Garc{\'i}a-P{\'e}rez,$^{1}$ Javier Ballesteros-Paredes$^{4}$, Marcial Becerril-Tapia$^{5}$
\\
$^{1}$Instituto Nacional de Astrof\'isica, \'Optica y Electr\'onica, Luis E. Erro 1, 72840 Tonantzintla, Puebla, M\'exico\\
$^{2}$Secretar\'{\i}a de Ciencia, Humanidades, Tecnolog\'{\i}a e Innovaci\'on (SECIHTI), Av. Insurgentes Sur 1582, 03940, Ciudad de M\'exico, M\'exico \\
$^{3}$Universidad Nacional Autónoma de M\'exico, Instituto de Astronom{\'i}a, A.P. 70-264, 04510, Ciudad de M\'exico, M\'exico \\
$^{4}$ Universidad Nacional Aut\'onoma de M\'exico, Instituto de Radioastronom\'ia y Astrof\'isica, Antigua Carretera a P\'atzcuaro 8701, Ex-Hda. San Jos\'e de la Huerta,\\ 58089, Morelia, Michoac\'an, M\'exico\\
$^{5}$Cardiff University, School of Physics and Astronomy, Cardiff, United Kingdom
}

\date{Accepted XXX. Received YYY; in original form ZZZ}

\pubyear{2026}

\begin{document}
\label{firstpage}
\pagerange{\pageref{firstpage}--\pageref{lastpage}}
\maketitle


\begin{abstract}
We investigated the morphology and strength of magnetic fields in pillar-shaped structures at the boundaries of \HII\ regions by combining three-dimensional radiation-magnetohydrodynamic (R-MHD) simulations with synthetic polarimetric and molecular-line observations. Our analysis focuses on a self-consistently formed pillar as a proof of concept to test the Davis-Chandrasekhar-Fermi (DCF) method under externally driven conditions. The pillar arises as an ionization front compresses a dense clump, producing a magnetically aligned, elongated structure whose morphology and field configuration resemble systems such as the pillars in M16. Synthetic 850~$\mu$m dust-polarization maps reproduce the pillar's large-scale magnetic-field morphology, confirming polarimetry as a reliable tracer of magnetic-field geometry. To evaluate DCF-based methods, we extract local density and velocity dispersion self-consistently from synthetic $^{13}$CO observations and measure polarization-angle dispersion using single-Gaussian fits to the synthetic polarization-angle distributions. We find that DCF-based methods systematically overestimate the intrinsic plane-of-sky magnetic-field strength by average factors of $\sim 7$ for the classical DCF method and $\sim 5$ for the modified Skalidis \& Tassis formulation. This overestimation is already present in the full-pillar measurement and is not removed by applying polarimetric S/N cuts or by excluding the dynamically complex head. We attribute the discrepancy to external compression by the expanding \HII\ region, which organizes the magnetic field on pillar scales while driving non-thermal gas motions. Consequently, the measured velocity and polarization-angle dispersions no longer trace the same turbulence-driven perturbation field assumed by DCF. Our results highlight the need for caution when applying DCF-based analyses to pillars or other externally compressed structures.

\end{abstract}




\defcitealias{Pattle2018}{P18}
\section{Introduction} \label{sec:introduction}

Interstellar structures known as pillars, elephant trunks, cometary globules, bright filaments, or bright rims are dense gaseous and dusty features with elongated shapes observed in \HII\ regions, planetary nebulae, supernova remnants (e.g., Crab Nebula), stellar wind shells (e. g. Bubble Nebula), and even interstellar cirrus  \citep[see ][]{jackson2003catalog}. Their origin is likely due to pre-existing inhomogeneities that are enhanced by radiative cooling in boundary layers \citep[see][]{hartquist1993clumps} and sculpted by dynamical \citep{giuliani1979hydrodynamic} and ionization-front instabilities \citep{vandervoort1962stability, williams2002instability, kim2014instability} in the flows of expanding \HII\ regions \citep{GarciaSegura1996ApJ, whalen2006multistep}, massive stellar winds \citep{garcia1996dynamical}, planetary nebulae \citep{garcia1999shaping}, and supernova remnants \citep{tenorio1991evolution, franco1991evolution}. Thus, they represent the structuring driven by mechanical and radiative feedback from stars, and their evolution is the result of the interplay of multiple physical processes. In particular, recent observations highlight the crucial role of magnetic fields in regulating their internal dynamics and maintaining long-term stability \cite[see e.g.,][]{Pattle2018, hwang2023}.

Only a few observational explorations have investigated the strength and configuration of magnetic fields in these dense and dusty structures, but these studies show important features. For example, polarimetric observations in the near infrared of the southern region of M16 show that the magnetic field in the ``Pillars of Creation'' is aligned with their main axes, whereas the field in the surrounding ionized region has different orientations \citep[e.g.,][]{Sugitani+2007}. This may indicate that strong UV radiation fields and associated gas flows affect the configuration and dynamical evolution of the magnetized dusty plasma. The M16 region is complex, with a number of cavities and neutral shells, where the parent molecular cloud has been shaped and eroded by the energy injected during different episodes of massive star formation \citep{karim2025sofiaM16}. Regarding the strength and structure of the B-fields, \citeauthor{Pattle2018} (\citeyear{Pattle2018}; hereafter \citetalias{Pattle2018}) analyzed the measurements of polarized dust emission at $850\,\mu\mathrm{m}$ with the aid of the classical Davis--Chandrasekhar-Fermi (DCF) method \citep[see Section \ref{sec:DCF};][]{davis1951, chandrasekhar1953}, and reported plane-of-sky magnetic field strengths of B$_\mathrm{\rm POS} \sim 170\text{--}320 \,\mu\text{G}$ for Pillar II, aligned with its main axis. These values were inferred for densities of $n \sim 5 \times 10^4\,\mathrm{cm}^{-3}$ and the velocity dispersions $\Delta v \sim \left(1.2\text{--}2.2\right)\,\mathrm{km\,s}^{-1}$ reported by \cite{White1999}, and \citetalias{Pattle2018} concluded that the magnetic field plays an important role in the stability and possible the longevity of the pillars. 

More recently, \cite{hwang2023} studied magnetic fields in the Horsehead Nebula using dust polarization measurements at $850\,\mu\mathrm{m}$. Two compact submillimetric sources in that region, SMM1 and SMM2, were analyzed with a modified version of the DCF method by \cite{Hildebrand2009}. In both sources, the magnetic field is oriented nearly perpendicular to their major axes. The derived plane-of-sky magnetic field strengths, for mean densities of up to $\sim 10^4\,\mathrm{cm}^{-3}$ and small velocity dispersions of $\sigma_v \approx 0.1\,\mathrm{km\,s}^{-1}$, were B$_\mathrm{\rm POS} \simeq 56 \pm 9\,\mu\mathrm{G}$ and $129 \pm 21\,\mu\mathrm{G}$ for SMM1 and SMM2, respectively. They concluded that these magnetic fields are strong enough to resist gravitational collapse in both sources, and the derived Alfvén Mach numbers indicate that magnetic pressure dominates over internal turbulent motions. Thus, the magnetic field strengths estimated using either the classical or the modified DCF method support a scenario in which an initially dynamically unimportant magnetic field permeating a structured molecular cloud is amplified by compressions from the expansion of neighboring \HII\ regions \citep[as observed in radiation–magnetohydrodynamic (R-MHD) numerical simulations, e.g.,][]{Bertoldi+1989, Williams07,Arthur2011}. 

The pressure of expanding photoionized regions compresses both the gas and the embedded magnetic fields, and the compressed magnetic field within the existing pillars can counteract further effects of such a pressure. This slows the evolution of the pillars by resisting lateral compressions \citep[e.g.,][]{Arthur2011} and prevents their shredding \citep{Tremblin+2012, Mackey+10}. Compressions of the B-field also reduce small-scale fragmentation, imposing large-scale order around the ionization front \citep[e.g.,][]{Arthur2011}. Then, the presence of magnetic fields significantly influences the dense gas around the \HII\ regions \citep[e.g.,][]{Krumholz2007ApJ, Henney+09}. \citet{Mackey_Lim2011} showed that moderate initial magnetic field strengths ($\simeq$ 50 $\mu$G) can produce pillars consistent with those observed in M16. This may also indicate that the large magnetic field strengths measured using the classical DCF method by \citetalias{Pattle2018} are likely due to the compression generated by the thermal and ram pressures of the expanding \HII\ region.

Numerical simulations have shown the influence of magnetic fields in pillar formation, but they also raise important questions regarding the interpretation of magnetic field strengths inferred from the classical DCF method. This widely used method assumes that the observed dispersion in the polarization angles arises from internal turbulent motions that perturb an otherwise uniform magnetic field \citep{chandrasekhar1953}. 

Early numerical tests by \citet{heitsch2001_DCFtestsInNumSims} revealed that observational limitations, such as limited telescope resolution and spatial subsampling, systematically overestimate magnetic field strengths by smoothing out small-scale fluctuations. They also demonstrated that the standard method fails in weak-field regimes where angular dispersions are large ($\delta\psi \sim 90^\circ$), necessitating corrections to constrain field strengths accurately.

Beyond these observational and numerical constraints, however, a physical discrepancy arises in the context of pillars. As stated above,  the field geometry in the pillars is likely dominated by externally driven compressions rather than internal turbulence, violating the energy-equilibrium assumption underlying the classical DCF approach. Thus, this classical method is not adequate under the conditions that prevail in the \HII\ regions. Our goal here is to explore the physics of magnetized pillars and assess the reliability of the classical DCF method in structures similar to those of these observed structures.

An additional physical ingredient that can play a role in the evolution of pillar-like structures is self-gravity. Only a few numerical studies have considered its influence, such as \citet{Gritschneder+2009} and \citet{Gritschneder2010} who showed that self-gravity amplifies inhomogeneities seeded by initial turbulence. However, these studies did not account for magnetic fields, highlighting the need for future models that include more physical processes because gravitational collapse can be triggered at the compressed and magnetized heads of the pillars, potentially leading to the formation of low-mass stars.

{We used 3D radiation-magnetohydrodynamic (R-MHD) simulations \citep[see][]{ZamoraAvils2019} coupled with synthetic polarimetric and molecular line observations to examine the magnetic field morphology and local gas kinematics. By extracting all the required physical parameters self-consistently and applying analytical techniques to decouple characteristic polarization-angle dispersion from large-scale field bending \citep[e.g.,][]{Palau2021, Polychronakis2025}, we evaluated the Davis-Chandrasekhar-Fermi (DCF) method. This approach allows us to quantify how the compression-driven magnetic field alignment on pillar scale affects the accuracy of magnetic field estimates and provides guidance for interpreting dust polarization observations of pillars in H \,{\sc ii} regions.}

This paper is organized as follows. The numerical setup for the R-MHD simulation and the generation of synthetic polarimetric and molecular-line maps are described in Section~\ref{sec:numerics}. Our main results, including the classical and modified DCF-based magnetic-field estimates, are presented in Section~\ref{sec:results}. In Section~\ref{sec:discussion}, we discuss the implications of these findings for the interpretation of magnetic-field observations in pillar-like structures. Finally, Section~\ref{sec:conclusions} summarizes the main conclusions and outlines the perspectives for future work. Additional methodological details are presented in the appendices: Appendix~\ref{sec:appendix_mass_estimation} describes the mass estimates from the synthetic molecular-line observations, Appendix~\ref{sec:appendix_secondOrderMomentmaps} shows the second-order moment maps, Appendix~\ref{sec:appendix_PolAngDist_gfits} presents the polarization-angle distributions and Gaussian fits, Appendix~\ref{sec:appendix_adf_analysis} describes the ADF analysis, and Appendix~\ref{sec:appendix_polaris_convergence} tests the robustness of the synthetic polarization morphology against the adopted POLARIS parameters.

\section{Numerical methods} \label{sec:numerics}

\subsection{Numerical model}\label{sec:Selfcons_Model}

We conducted a synthetic polarimetric study of a pillar-like structure extracted from the RMHD simulation of \cite{ZamoraAvils2019} performed with the AMR FLASH code \citep[v2.5][]{FLASH1}, whose objective was to study the structure and evolution of \HII\ regions. This simulation follows the full evolution of a molecular cloud (MC) formed through compressions caused by converging flows in the warm neutral medium (WNM) and includes the relevant physical process for studying magnetic fields in photoionized regions. These processes comprise magnetic fields \citep{Waagan+11}, self-gravity implemented via the OctTree algorithm \citep{Wunsch+18}, heating and cooling \citep{KI00, KI02, Peters_2010}, sink particle formation \citep{Federrath+10}, and ionizing feedback from massive stars using a ray-tracing scheme \citep{Rijkhorst06,Peters_2010}. 

The initial conditions and numerical setup are as follows. The simulation box has dimensions of $256\times128\times128\,\mathrm{pc}^3$ and contains a warm neutral gas with a density of $n=2\,\mathrm{cm}^{-3}$ and a temperature of $T=1450\,\mathrm{K}$, permeated by a uniform magnetic field of $3\umuG$ oriented along the $x$ axis. Two cylindrical WNM streams, each 112 pc in length and 32 pc in radius, collide head-on along the $x$-axis at the center of the box with a velocity of $7.5\ \ukms$, forming a compressed layer. A background subsonic turbulence with Mach number 0.7 is imposed to break the layer symmetry and to trigger dynamical instabilities. The simulation grid dynamically refines according to the Jeans criterion \citep{Truelove+97}, achieving a maximum resolution of $0.03\,\mathrm{pc}$ in the densest regions.

Thermal instability \citep{Field_65} drives a phase transition in the shocked layer from the WNM to the cold neutral phase. Around $11.6\,\mathrm{Myr}$, the layer becomes dense enough to become molecular and gravitationally unstable \citep[see e.g.][]{Vazquez_Semadeni+19}. The densest clumps undergo rapid collapse, eventually forming massive stars that begin to photoionize their surrounding clumpy medium, creating pillar-shaped structures through compression and erosion processes at the external boundaries of the resulting \HII\ regions.

\subsection{Pillar identification} \label{subsec:PillarIdentification}

The pillars in M16 are characterized as dense ($n \gtrsim 10^4\,\mathrm{cm^{-3}}$), elongated structures with typical dimensions of $\sim 1$ pc along their main axis and $\sim 0.2$ pc along the minor axis \cite[see e.g. \citetalias{Pattle2018};][]{Hester1996, hwang2023}. To enable a consistent comparison with observations, we adopt the following procedure to identify and track pillar-like structures in the simulation. First, we locate overdense regions surrounding the \HII\ regions. We then select cells that satisfy $n \gtrsim 300\,\mathrm{cm^{-3}}$, ionization fraction $\ll 1$, and $T < 30$ K, located within a sphere of radius $R_{\text{thr}} \leq 1$ pc centered on the local density peak \cite[e.g.][]{Walch2012, pattle2022magnetic}.

For the selected dense cold gas cells, we compute the three principal axes using the inertia matrix, $\mathbf{I}$, with components defined as:
\begin{equation}
    \text{I}_{ij} = \sum_k \Delta m_k \left( \| \mathbf{r}_k \|^2 \delta_{ij} - x_i^{(k)} x_j^{(k)} \right),
\end{equation}
where $\Delta m_k$ is the mass of the $k$-th cell; $\mathbf{r}_k=\left(x_1^{(k)}, x_2^{(k)}, x_3^{(k)} \right)$ is the position vector of the cell relative to the center of mass; and $\delta_{ij}$ is the Kronecker delta. We then calculate the eigenvalues of the inertia matrix, $I_i$ (representing the principal moments of inertia), and estimate the semi-axes assuming an ellipsoidal geometry. The major semi-axis is given by $a=\sqrt{5(I_2+I_3-I_1)/2M}$ and the minor semi-axis by $c=\sqrt{5(I_1+I_2-I_3)/2M}$, where $M$ is the total mass of the dense gas structure. Note that $a$ and $c$ are dynamically defined at each timestep as the instantaneous maximum and minimum semi-axes, respectively.

\begin{figure*}
    \centering
    \includegraphics[width=\textwidth]{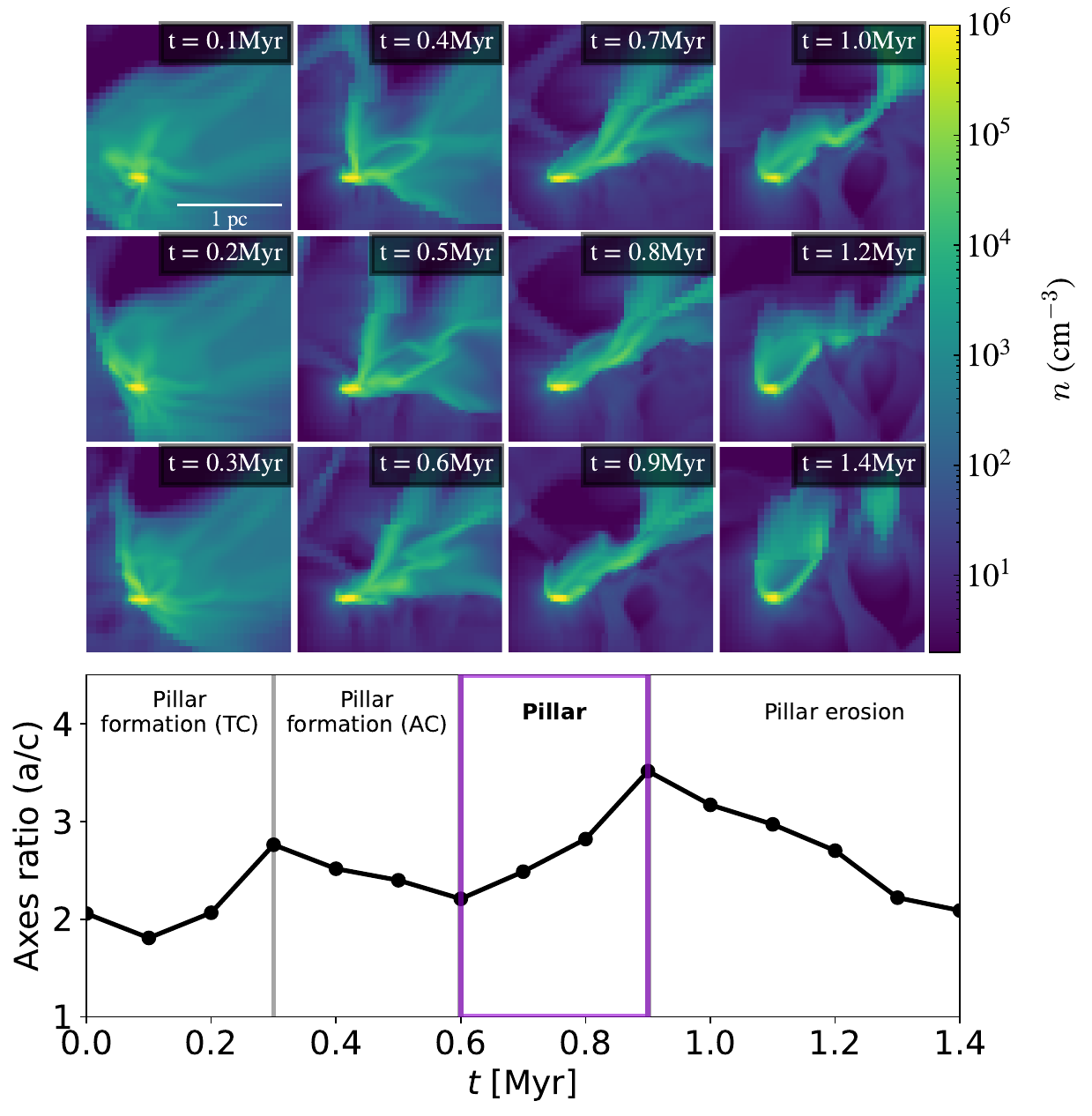}
    \caption{ {Top panel:}Representative gas density slices in the $z$–$y$ plane are shown to visually illustrate the morphological evolutionary stages of the structure at corresponding times. {Bottom panel:}  Time evolution of the ratio between the major and minor axes ($a/c$) of the dense gas structure, computed from the inertia matrix. { As described in Sec.\ref{subsec:PillarIdentification}, the distinct regions of the plot delimit the evolutionary stages of the structure: transverse compression (TC), the phase of pillar formation via axial compression (AC), the stable pillar lifetime (column with the label ``Pillar'' and highlighted in purple), and its destruction phase. The axes ratio traces the full morphological evolution of the structure, increasing as the pillar becomes more elongated during its stable phase and then decreasing as erosion progressively disrupts it.} }
    \label{fig:pillar_evol_ac_ratio}
\end{figure*}

Figure \ref{fig:pillar_evol_ac_ratio} illustrates the time evolution of the aspect ratio ($a/c$) measured for the dense gas ({bottom panel}, solid black line). The vertical lines show the distinct evolutionary stages, with the stable pillar lifetime highlighted in purple. Additionally, representative snapshots of the evolution of the pillar are shown as $2 \times 2$ pc$^2$ density slices in the $y$--$z$ plane ({top panel}). These snapshots are shifted relative to the density peak to capture the full extent of the gas, illustrating the morphological changes corresponding to the Transverse Compression, Axial Compression, Pillar, and Pillar erosion phases, respectively. Tracking the behavior of the $a/c$ ratio allows us to identify four evolutionary stages:

1. Transverse Compression (TC): This stage is characterized by the initial flattening of the structure as the shock front impacts the over-density. The compression acts primarily along the direction of shock propagation, which corresponds to the minor axis $c$ at this stage. Consequently, $c$ decreases while $a$ remains relatively constant or decreases at a slower rate, resulting in a sharp increase of the aspect ratio $a/c$ (see the ``Pillar formation (TC)'' column in Fig.~\ref{fig:pillar_evol_ac_ratio}, { bottom panel}).

2. Axial Compression (AC): After initial flattening, the shock front envelopes the structure. During this phase, the aspect ratio $a/c$ decreases over time. Physically, this corresponds to a "squashing" of the structure's major extent as the shock sweeps up and compresses the gas from the sides. It is important to clarify that because $a$ and $c$ are defined by instantaneous geometry, the physical axes may change roles during this transition: the axis aligned with the flow (initially compressed as $c$ during TC) can eventually evolve to become the main axis $a$ of the resulting pillar. The AC phase represents the geometric transition where the longest dimension is significantly reduced relative to the thickness (see the ``Pillar formation (AC)'' column in Fig.~\ref{fig:pillar_evol_ac_ratio}, { bottom panel}). 

3. Life span of the pillar: The pillar is considered fully formed once the $a/c$ ratio begins to increase again (at $t \approx 0.6$ Myr). At this point, the structure has achieved an elongated geometry where erosion along the flanks is balanced by external thermal pressure (see the ``Pillar'' column in Fig.~\ref{fig:pillar_evol_ac_ratio}, { bottom panel}).

4. Pillar erosion: The lifetime of the pillar ends when the structure begins to disconnect from its parent cloud (see the column `` Pillar erosion" in Fig.~\ref{fig:pillar_evol_ac_ratio}, { bottom panel}).

Regarding the duration of these stages, the AC and Pillar phases are strictly defined by the inflection points in the evolution of the $a/c$ ratio. In contrast, the onset of the TC phase and the end of the Pillar erosion phase are visually determined based on morphology. Therefore, while the core formation and lifetime phases are mathematically rigorous, the start and end points of the entire process are somewhat arbitrary in duration. Furthermore, since these measurements are constrained by the temporal resolution of the simulation outputs, the reported values should be considered representative estimates rather than exact timings.

\begin{figure*}
    \centering
    \includegraphics[width=\textwidth]{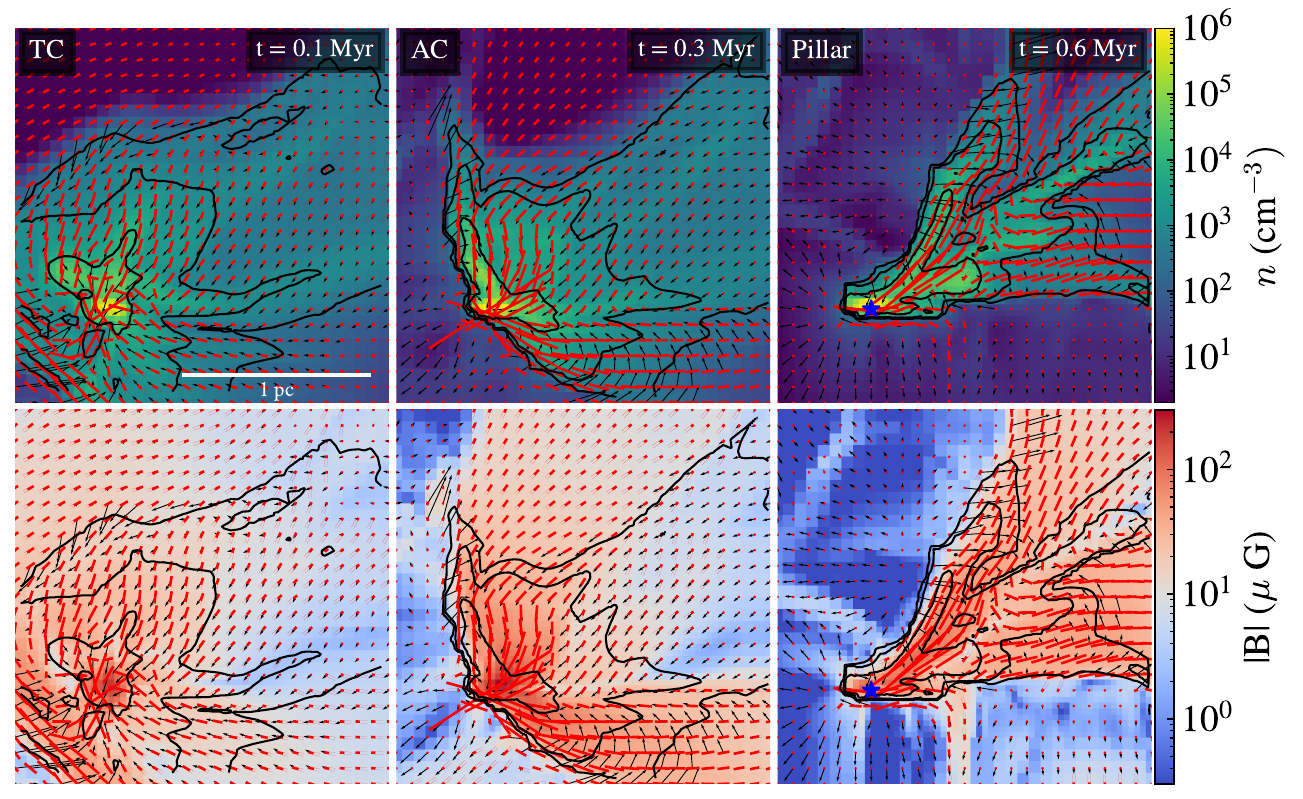}
    \caption{Slices in the $z$-$y$ plane illustrating the formation process of the first pillar in the simulation. The upper and lower panels show the number density and the magnetic field strength at three different times. The corresponding evolutionary stages (TC, AC, and Pillar) are indicated by labels in the upper left corner of the panels. Black contours indicate number density levels of $3 \times 10^2, \, 1 \times 10^3 \, \rm{and} \, 1 \times 10^4 \uicmc$. Black arrows depict the velocity field, while red line segments indicate the magnetic field orientation and relative strength. The blue star marks the position of one of the sources embedded within the pillar tip (symbol size not to scale). { The magnetic field is progressively bent and aligned along the pillar as external compression strengthens during its formation.} } 
    \label{fig:multiplot_mdlAutoC_2x3}
\end{figure*}

\subsection{The Davis–Chandrasekhar–Fermi (DCF) method} \label{sec:DCF}

As stated above, one of the most widely used techniques to estimate the magnetic field strengths from polarized dust emission is the classical DCF method \citep{davis1951, chandrasekhar1953}. This method provides an estimate of the projected plane-of-sky magnetic field strength $B_{\rm POS}$ in a given dusty gas structure, by relating internal turbulent motions to perturbations in the magnetic field. It is based on three key assumptions: {\it (i)}~the gas is incompressible, {\it (ii)}~there is an energy balance between turbulent kinetic fluctuations ($\delta E_{K}$) and magnetic fluctuations ($\delta E_{B}$), and {\it (iii)}~the turbulence is isotropic in three dimensions. 

Consider a magnetized medium with a uniform background magnetic field $B_0$ and a perturbation $B_{\perp}$, induced by a turbulent velocity fluctuation perpendicular to the field, $v_{\perp}$ \citep[see, e.g.,][]{Chen2022}. Under assumption {\it (ii)}, the energy balance can be written as
\begin{equation}
\frac{\delta E_{K}}{\delta E_{B}} = \frac{\tfrac{1}{2}\rho v_{\perp}^{2}}{B_{\perp}^{2}/(8\pi)} = 1,
\end{equation}
where $\rho$ is the density of the gas. This implies a direct proportionality between turbulent and magnetic perturbations, mediated by the Alfvén velocity ($V_{A} \equiv \sqrt{B_0^2/4 \pi \rho}$), such that $v_{\perp} = V_{A}(B_{\perp}/B_{0})$. From this, the uniform field strength can be written as
\begin{equation}
    B_0 = \sqrt{4\pi \rho}\,\frac{v_{\perp}}{B_{\perp}/B_0}.
\end{equation}

In practice, $B_{\rm POS}$ can be derived from observational quantities. The angular dispersion of the polarization segments, $\delta \psi$, serves as a proxy for $B_{\perp}/B_0$, while the turbulent velocity is estimated from the non-thermal velocity dispersion of the gas, $\sigma_v$, which is related to the FWHM velocity width through $\Delta v = \sigma_v \sqrt{8 \ln 2}$. This leads to
\begin{equation}\label{eq_Bpos_clas_f05} 
   B_{\text{\rm POS}} \approx f \sqrt{4 \pi \rho}\,\frac{\sigma_v}{\delta \psi}, 
\end{equation} 
where $f$ is a correction factor that accounts for the effects of beam averaging and line-of-sight integration. The commonly adopted value $f=0.5$ was calibrated using MC simulations \citep{Ostriker2001}. However, under the magnetic-turbulent energy equipartition conditions and within the validity of the small-angle approximation ($\delta \psi < 25^{\circ}$), the 3D MHD simulations of \citet{liu2021calibrating} showed that a lower value of $f \sim 0.25$ is more appropriate for dense clumps and cores ($<0.1$ pc).

A strong limitation of the classical DCF approach is that it assumes incompressibility. In reality, the interstellar medium is highly compressible, and turbulent motions include not only Alfvénic modes, but also fast and slow magnetosonic modes. These additional modes contribute to velocity dispersions and broaden polarization angle distributions, causing $\sigma_{v,turb}$ to exceed the Alfvénic contribution alone. As a result, the classical DCF method systematically overestimates $B_0$ \citep{Skalidis2021}. 

To address this, \citet{Skalidis2021sqrtofdPA} proposed a modified formulation that relaxes the assumption of incompressibility. In the sub- or trans-Alfvénic regime, the term $\vec{B}\cdot\delta\vec{B}\neq 0$ dominates the magnetic energy density and can be equated with the kinetic energy of turbulent motions ($\rho \delta v^2/2 = \delta B\,B_0/4\pi$). This leads to the modified DCF relation:
\begin{equation} \label{eq:DCFmodifiedS21}
    B_0 \approx \sqrt{4\pi \rho}\,\frac{\delta v}{\sqrt{2\delta \psi}},
\end{equation}
which is used in this work and denoted by $B_{\rm POS,ST21}$.

\subsection{Synthetic observations}\label{subsec:so}

In order to compare the magnetic field strength from our simulations with the estimates obtained using the mentioned classical and modified DCF methods, we post-process our numerical data through radiative transfer (RT) synthetic observations. This allows us to derive the required observational parameters: the mean density ($\rho$), the velocity dispersion ($\sigma_v$), and the angular dispersion of the polarization vectors ($\delta \psi$).

\subsubsection{Dust polarization maps}

To calculate the orientation of the magnetic field, we use the \texttt{POLARIS}\footnote{\url{http://www1.astrophysik.uni-kiel.de/~polaris}} code \citep{Reissl2016}, which simulates dust continuum emission and dust polarization using the Monte Carlo technique, taking into account the relevant physical processes responsible for dust alignment. 

Following the study of 3D MHD polarization simulations by \citet{Seifried+19}, we adopt typical parameters for dusty molecular clouds as input for \texttt{POLARIS}. The dust-to-gas mass ratio in nearby galaxies depends on metallicity \citep[][]{remy2014dustgasmassratios, roman2022dust2gasratio_metal}, and the integrated value for the Milky Way is about $8\times 10^{-3}$ \citep[see][and references therein]{franco2025pressure_dust2gasmssratio}. For simplicity, here we use $10^{-2}$ and assume a dust grain mixture of $62.5\%$ silicates and $37.5\%$ graphite, with a size distribution $n_d(a) \propto a^{-3.5}$, where $n_d$ is the volume density of the dust and $a$ is the effective grain radius, ranging from $a_{\mathrm{min}} = 5 \times 10^{-9}\,\mathrm{m}$ to $a_{\mathrm{max}} = 250 \times 10^{-9}\,\mathrm{m}$ \citep{Mathis1977, Draine2001}. We employ the \texttt{POLARIS} Radiative Alignment Torques (RAT) module \citep{Hoang2009} to compute the dust temperature $T_d$ and the alignment radius $a_{\mathrm{alg}}$, with a fraction of grains of the high-$J$ state of $f_{\mathrm{high-}J} = 0.25$. 

The flux of polarized light is measured at $850\,\mu\mathrm{m}$, consistent with \citetalias{Pattle2018}. Linear polarization is then derived from the Stokes parameters $I$, $Q$, and $U$, with the polarization angle $\psi$ calculated as \citep{Reissl2016}:

\begin{equation}
    \psi = \frac{1}{2} \arctan\left(\frac{U}{Q}\right)
    \label{eq:polarisation_angle}
\end{equation}
and the degree of linear polarization $p$ as:
\begin{equation}
    p = \frac{\sqrt{U^2 + Q^2}}{I}.
    \label{eq_linpolfrac}
\end{equation}

The dust heating sources are three stars that self-consistently formed within the simulation. The most massive star, primarily responsible for generating the \HII\ region, has an effective temperature of $T_{\mathrm{eff}} \approx 21{,}127\,\mathrm{K}$ and an ionizing photon rate of $Q_{\mathrm{H}} \sim 10^{45}\,\mathrm{s^{-1}}$, and is located approximately $2.52\,\mathrm{pc}$ from the tip of the pillar. A second star, with $T_{\mathrm{eff}} \approx 19{,}705\,\mathrm{K}$, lies about $3.58\,\mathrm{pc}$ from the primary star, while a third star, with $T_{\mathrm{eff}} \approx 14{,}433\,\mathrm{K}$, is embedded within the tip of the pillar (indicated by the blue star in the right panels of Figure~\ref{fig:multiplot_mdlAutoC_2x3}). Following \citet{Mathis_1983} and \citet{Camps_2015}, an interstellar background radiation field is also included. 

{  To characterize the polarimetric signal-to-noise (S/N) ratio in the synthetic observations, we compute the ratio $p/\delta p$ directly from the Stokes parameters.} The linear polarized intensity is defined as
$I_L = \sqrt{Q^2 + U^2}$,
and the linear polarization fraction is
$p = \frac{I_L}{I}$.
Assuming independent uncertainties in the Stokes parameters, the uncertainty in $I_L$ is obtained by standard error propagation as
\begin{equation} 
    \delta I_L =
    \frac{1}{I_L}
    \left[
        (Q\,\delta Q)^2 + (U\,\delta U)^2
    \right]^{1/2}.
\end{equation}

{  We estimate the Stokes uncertainties from the rms levels measured in a low-column-density background region of the projected maps. This region comprises lines of sight outside the pillar footprint that are dominated by diffuse, warm, predominantly ionized gas, thereby excluding the dense, cold, predominantly neutral columns used to identify the pillar (see Section~\ref{subsec:PillarIdentification}).}

The same background selection is applied consistently to $I$, $Q$, and $U$. For a given Stokes parameter $X\in\{I,Q,U\}$, the rms value is computed as
\begin{equation} 
    X_{\mathrm{rms}} =
    \left(
        \frac{1}{N_{\mathrm{bg}}}
        \sum_{i=1}^{N_{\mathrm{bg}}} X_i^2
    \right)^{1/2},
\end{equation}
where $N_{\mathrm{bg}}$ is the number of pixels in the selected background region. We then set
$\delta I \equiv I_{\mathrm{rms}}$,
$\delta Q \equiv Q_{\mathrm{rms}}$, 
$\delta U \equiv U_{\mathrm{rms}}$.

The uncertainty in the polarization fraction is therefore
\begin{equation}
    \delta p =
    \left[
        \frac{\delta I_L^2}{I^2}
        +
        \frac{I_L^2\,\delta I^2}{I^4}
    \right]^{1/2}.
\end{equation}
This allows us to construct the polarimetric S/N map, $p/\delta p$, as well as the total-intensity S/N map, $I/\delta I$.

{  In our main analysis, the fiducial configuration corresponds to the full projected pillar without applying polarimetric S/N masks, which is the configuration most directly comparable to the physical pillar identified in the simulation. The strict cuts $I/\delta I>10$ and $p/\delta p>3$, following the observational criterion adopted by \citetalias{Pattle2018}, are instead used to define comparison samples. These S/N-selected full-pillar and body-only measurements allow us to test whether the inferred angular dispersions and DCF-based magnetic-field estimates depend on the adopted polarimetric selection.}

\subsubsection{Molecular line observations}

To derive the kinematic and column density properties of the gas, we performed complementary molecular-line radiative-transfer (RT) calculations using the \texttt{LIME}\footnote{\url{https://github.com/lime-rt/lime/tree/master}} code \citep{Brinch2010} to generate synthetic observations of the $^{13}\mathrm{CO}(J=1-0)$ transition lines. To ensure full consistency between the two radiative transfer calculations, we extract from the simulation exactly the same region used in \texttt{POLARIS} as input for \texttt{LIME}. We likewise use the same synthetic detector resolution, $256 \times 256$ pixels. We assumed a uniform $^{13}\mathrm{CO}$ abundance of $1.3 \times 10^{-6}$ relative to $\mathrm{H}_2$ \citep[see, e.g.,][]{Sofue2020} and included an isotropic microturbulent velocity characterized by a Doppler broadening parameter of $100\,\mathrm{m\,s^{-1}}$. The resulting Position-Position-Velocity (PPV) data cubes were projected at a simulated distance of $1\,\mathrm{kpc}$, spanning $100$ channels with a velocity resolution of $0.1\,\mathrm{km\,s^{-1}}$ per channel. This value was adopted from the literature to maintain consistency with high-resolution observations of the pillars in M16 \citep[e.g.,][]{karim2023sofiaPillars}.

\subsubsection{Parameter extraction for the DCF method} \label{sec:parameters}

To determine the magnetic field strength using DCF-based methods (Eqs. \ref{eq_Bpos_clas_f05} and \ref{eq:DCFmodifiedS21}), we extract all relevant parameters directly from our synthetic observations as follows:

\begin{enumerate}[(i)]
    \item {  {Density ($\langle \rho \rangle_{\mathrm{cyl}}$).} We used Synthetic Observation Software (\texttt{SOS}\footnote{\url{https://github.com/MarcialX/sos/tree/master}}) to estimate the total mass of the pillar using the conversion factor (X-factor) method ($M_{\mathrm{XF}}$) applied to our synthetic maps of $^{13}$CO (see Appendix~\ref{sec:appendix_mass_estimation}).} 
    
    Following the standard observational procedure, in which an assumed 3D geometry is adopted to infer the mean mass density from the integrated $^{13}$CO intensity map, we model the pillar as an effective cylinder of volume $V_{\rm cyl}$, with its longitudinal axis aligned with the major axis of the pillar. We then define the mean volumetric density as $\langle \rho \rangle_{\mathrm{cyl}} = M_{\rm XF}/V_{\rm cyl}$. This provides a self-consistent estimate of the mean volumetric density at each time (see Section~\ref{subsec:mass_density}).
    
    \item {Velocity dispersion ($\sigma_v$). The representative non-thermal line width ($\Delta v$) was extracted from the second spectral moment map of the synthetic $^{13}$CO emission (see Appendix~\ref{sec:appendix_secondOrderMomentmaps}).  For the fiducial analysis, we measure $\Delta v$ over the full projected pillar without applying the polarimetric S/N masks. The local velocity dispersion entering the DCF-based estimates is then obtained as
    $\sigma_v = \Delta v / \sqrt{8 \ln 2}$ 
    (see Section~\ref{subsec:velocity_dispersion}).We also compute $\Delta v$ for the S/N-selected full pillar and for the S/N-selected pillar body. These alternative selections give values that are not significantly different from the fiducial full-pillar measurement.}

    \item {Polarization-angle dispersion ($\delta\psi_{\mathrm{gfit}}$). The polarization-angle dispersion was measured from the synthetic \texttt{POLARIS} maps using a Gaussian fitting analysis of the polarization-angle distributions. In the fiducial configuration, the distribution is constructed from the full projected pillar without applying the polarimetric S/N masks. The S/N-selected full-pillar and body-only S/N-selected regions are analyzed separately to test the sensitivity of the inferred dispersion to data-quality and spatial-selection effects.}

    The fitting procedure allows for multiple Gaussian components when statistically justified, following the motivation of \citet{Palau2021} and \citet{Polychronakis2025}. However, in our analysis, nearly all timesteps are well described by a single Gaussian component. We therefore adopt the single-Gaussian width as our main estimate of the characteristic angular dispersion, $\delta\psi_{\mathrm{gfit}}$ (see Section~\ref{subsec:SPObs_pillar} and Appendix~\ref{sec:appendix_PolAngDist_gfits}).

    As an independent comparison, we also compute the Angular Dispersion Function (ADF; \citealt{Hildebrand2009}). The ADF results are shown for the full projected pillar, both with and without polarimetric S/N masking (see Appendix~\ref{sec:appendix_adf_analysis}). These ADF measurements are used solely as a consistency check on the angular dispersion obtained with the Gaussian fitting method. They are not used to compute $B_{\mathrm{POS}}$ in any of the DCF or ST21 magnetic-field estimates. Throughout the main analysis, the magnetic-field strengths are computed using $\delta\psi_{\mathrm{gfit}}$.

\end{enumerate}

By substituting these self-consistently derived parameters ($\langle \rho \rangle_{\mathrm{cyl}}$, $\sigma_v$, and $\delta\psi_{\mathrm{gfit}}$) into Eqs.~\ref{eq_Bpos_clas_f05} and~\ref{eq:DCFmodifiedS21}, we obtain synthetic DCF-based estimates of the plane-of-sky magnetic field strength ($B_{\mathrm{POS}}$). {  These estimates depend entirely on the physical and observational properties extracted from the simulated pillar, allowing us to directly compare the DCF-inferred field with the true magnetic field strength measured from the 3D R-MHD simulation.}

\begin{figure}
    \centering
    \includegraphics[width=\linewidth]{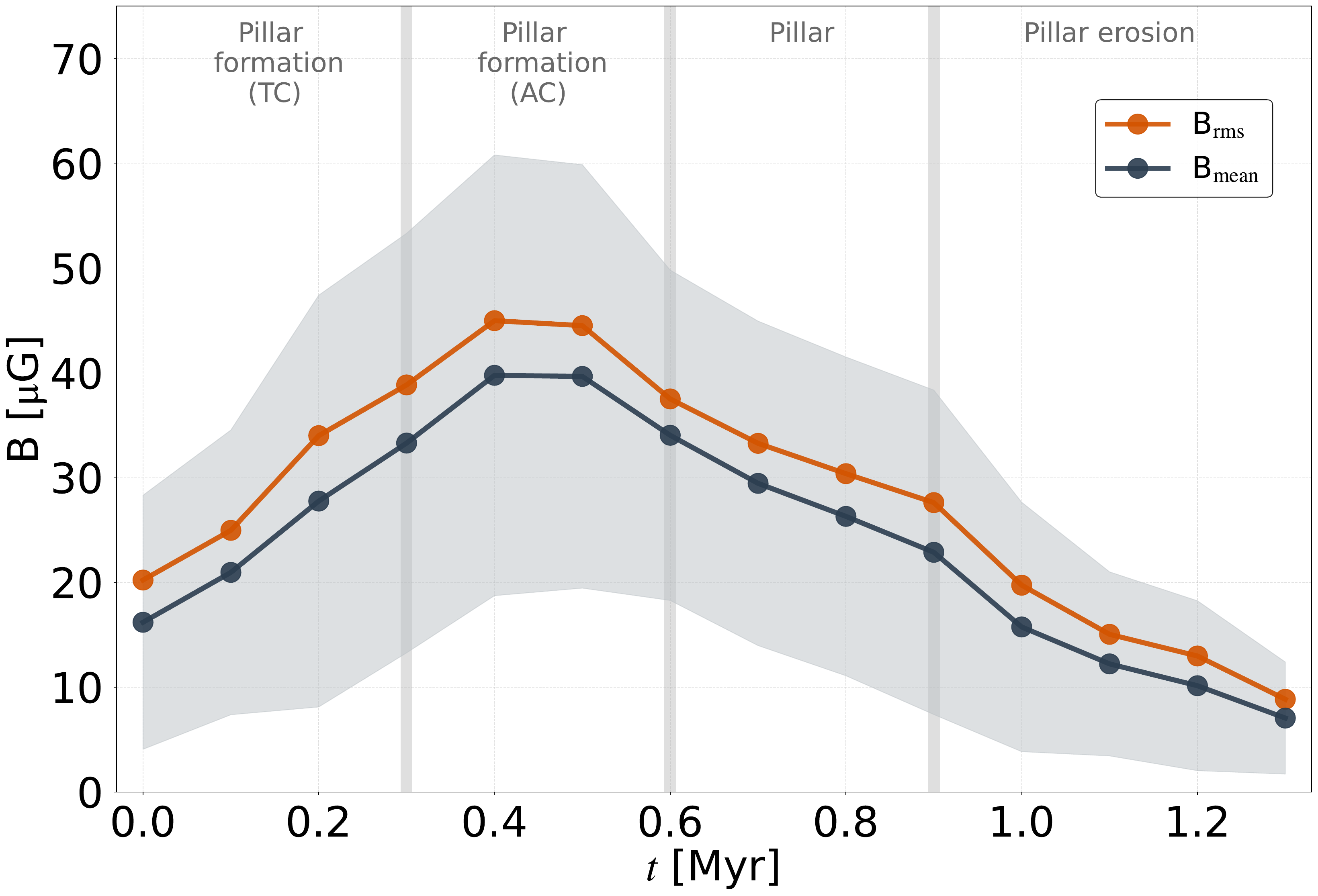}
    \caption{Evolution of the mean (black line, the gray shaded region indicates its standard deviation) and rms (orange line) magnetic field strengths in the dense gas $n \geq 300\,\mathrm{cm^{-3}}$ over time. The rms magnetic field strengths in the pillar stage ($t \sim 0.6-0.9$ Myr) range from 38 to 27 $\mu$G. The vertical lines delimit approximately the phases of pillar formation, evolution, and erosion, as in Fig.~\ref{fig:pillar_evol_ac_ratio}. The magnetic field is amplified during the pillar-formation phase, peaks near the onset of the stable pillar stage, and then gradually declines as the pillar evolves and erodes.} 
    \label{fig:Bfield}
\end{figure}

\section{Results}\label{sec:results}

For a consistent comparison between the 3D simulations and the 2D synthetic observations, we restricted both analyses to the same subset of cells (see Section \ref{subsec:PillarIdentification}). To measure the physical properties of the dense gas and perform radiative transfer simulations, we considered only cells with densities above the threshold value $n_{\rm thr} \geq 300 \, \mathrm{cm^{-3}}$. In addition, we focus on the first pillar that forms in the simulation as a proof of concept (see Fig. \ref{fig:multiplot_mdlAutoC_2x3}). For clarity in our discussion, we define time zero as approximately the moment when the ionization front reaches the dense region under study.

\subsection{Qualitative description of the pillar formation} \label{sec:pillar_formation}

The pillar formation process is driven by the interaction between the ionization and shock fronts of the expanding \HII\ region and the back-reaction of the pre-existing magnetized clump. This process takes approximately $t \lesssim 0.5$ Myr after the ionization front reaches the dense gas structure, and its evolution is quantitatively tracked in Fig. \ref{fig:pillar_evol_ac_ratio}.

In the early stage of formation (TC, see Section \ref{subsec:PillarIdentification} and Fig. \ref{fig:pillar_evol_ac_ratio}), the clump is compressed by the sweeping effect of the expanding \HII\ region, eventually becoming the head of the proto-pillar. This compression results in the initial flattening of the structure, visible as the sharp increase in the $a/c$ ratio in Fig.~\ref{fig:pillar_evol_ac_ratio}. As the ionization front advances, the structure transitions to a second stage of formation (AC), characterized by a configuration elongated perpendicular to the direction of the UV source. The compression acting on this configuration causes the $a/c$ ratio to decrease (Fig.~\ref{fig:pillar_evol_ac_ratio}). During this compression, the magnetic field (initially dynamically unimportant) is squeezed and amplified. As shown in Fig. \ref{fig:Bfield}, both the mean magnetic field strengths ($B_{\rm mean}$) and root-mean-square ($B_{\rm rms}$) increase significantly, reaching their peak values as the structure transitions into the next phase. The head protects the gas behind it from stellar UV radiation, and the nascent pillar becomes aligned along the direction of the UV source, consistent with the non-magnetic simulations of \citet{GarciaSegura1996ApJ}. The combined action of the ram pressure and the ionization front squeezes and stretches the gas, progressively elongating the pillar structure. The resulting cometary shapes are similar to those described by \citet{Arthur2011} in their simulations of turbulent and magnetized \HII\ regions, as well as those found in the interaction of high-velocity clouds with a magnetized galactic halo \citep{santillan1999collisions}, where erosion from the ram pressure and the tension of the compressed magnetic field drive the evolution of the head-tail configuration.

Finally, the pillar enters the erosion phase (see Fig. \ref{fig:pillar_evol_ac_ratio}). As external thermal pressure gradually weakens \citep{ZamoraAvils2019}, the head begins to relax (indicated by a decrease in the ratio $a/c$) and eventually decouples from the tail. During this phase, the mean density and the magnetic field intensity of the pillar decrease, as shown in Fig. \ref{fig:Bfield}. This evolutionary picture is consistent with the findings of previous works \citep[e.g.,][]{Bertoldi+1989, Gritschneder2010, Mackey_Lim2011, Arthur2011}.

A spatial view of this process is presented in Fig. \ref{fig:multiplot_mdlAutoC_2x3}, which displays slices in the $z$--$y$ plane at three representative times ($t=0.1, 0.3$, and $0.6$ Myr). At $t=0.1$ Myr (left panels), corresponding to the TC phase, the density contours show the clump flattened by the shock. The magnetic field vectors (red segments) begin to bend but generally maintain their initial orientation. At $t=0.3$ Myr (middle panels), during the AC phase, the structure adopts a cometary shape. The velocity field (black arrows) shows the ionized gas flowing past the clump, whereas within the pillar, the flow is directed toward the overdensity. At this stage, the magnetic field strength (lower panel color map) reaches its maximum in the compressed head. The pillar reaches its final elongated form after the AC phase at $t=0.6$ Myr (right panels). Notably, the magnetic field lines in the tail are well-ordered and aligned parallel to the pillar's main axis. The structure remains stable during this phase, and the magnetic field provides a stabilizing force that helps maintain its shape.

\subsection{Physical properties of the pillar}

The dense gas defining the pillar (cells with $n>10^3$ cm$^{-3}$, see Section \ref{subsec:PillarIdentification}) has a lifetime of approximately $0.3$~Myr and the simulation displays physical properties consistent with those reported for the pillars in M16 at a time $t = 0.7$~Myr after the formation of the pillar (see Table \ref{tab:pillar_comparison}).

\begin{table}
    \centering
    \caption{Comparison between the physical parameters of the simulated pillar at $t = 0.7$~Myr and representative observational values for Pillar~II in M16.}
    \label{tab:pillar_comparison}
    \setlength{\tabcolsep}{3pt} 
    \begin{tabular}{l c c r}
        \hline
        Parameter & Simulated & Observed & Ref. \\
        \hline
        Lifetime (Myr) & $< 0.3$ & $0.1$--$3$ & [1, 2, 6] \\
        Mass ($M_{\odot}$) & $\sim 99.3$ & $\sim 56$\textsuperscript{a}, $103$\textsuperscript{b} & [3, 6] \\
        Mean density (cm$^{-3}$) & $\sim 1.8 \times 10^4$ & $(2$--$5) \times 10^4$\textsuperscript{c} & [3, 4, 6] \\
        Radius (pc) & $\sim 0.33$ & $\sim 0.15$ & [4, 6] \\
        Length (pc) & $\sim 0.98$ & $1$--$1.5$\textsuperscript{d} & [5] \\
        Line width $\Delta v$ (km s$^{-1}$) 
        & $  {1.53\pm0.62}$\textsuperscript{e} 
        & $1.2$--$2.2$ 
        & [3, 4] \\
        Dispersion of & $  {10.9\pm0.8^{\circ}}$\textsuperscript{f} & $14.4^{\circ}$ & [4] \\
        polarization angles ($\delta \psi$) & & & \\
        Magnetic-field  
        & $  {33.8}$\textsuperscript{g}, 
          $  {279\pm115}$\textsuperscript{h}, 
           
        & $170$--$320$ 
        & [4] \\
        strength ($\mu$G) & $  {172\pm70}$\textsuperscript{i} & & \\
        \hline
    \end{tabular}

    \begin{flushleft}
        \footnotesize{
        {Notes:} Observational values are representative. {  The synthetic $\Delta v$, $\delta\psi$, and $B_{\rm POS}$ values correspond to the fiducial configuration (full-pillar and S/N unmasked).} \\
        \textsuperscript{a} Mass of H$_2$ derived from the C$^{18}$O map. \\
        \textsuperscript{b} Total mass derived from atomic gas mass and molecular gas mass. \\
        \textsuperscript{c} Range from body ($2\times10^4$) to global ($5\times10^4$) averages. \\
        \textsuperscript{d} Approximate values estimated visually from the scale bars in Figs.~1 and~2 of [5]. \\
        \textsuperscript{e} Spatially averaged synthetic $^{13}$CO$(J=1-0)$ line width, measured from the second spectral moment. \\
        \textsuperscript{f} Polarization-angle dispersion measured from the single-Gaussian fit to the full projected pillar. \\
        \textsuperscript{g} $B_{\rm rms,3D}$ measured directly from the simulation. \\
        \textsuperscript{h} $B_{\rm POS}$ derived using the classical DCF method. \\
        \textsuperscript{i} $B_{\rm POS,ST21}$ derived using the modified DCF method of \citet{Skalidis2021sqrtofdPA}. \\
        {References:} [1]~\citet{Williams+2001}; [2]~\citet{McLeod2015}; [3]~\citet{White1999}; [4]~\citetalias{Pattle2018}; [5]~\citet{Hester1996}; [6]~\citet{karim2023sofiaPillars}.
        }
    \end{flushleft}
\end{table}


\subsection{Local non-thermal velocity dispersion}
\label{subsec:velocity_dispersion}

To accurately estimate the magnetic field strength using DCF-based methods, it is  important to properly determine the local non-thermal velocity dispersion of the gas. As described in Section~\ref{subsec:so}, we derived the line width ($\Delta v$) and the corresponding velocity dispersion ($\sigma_v$) from the second spectral moment of the synthetic $^{13}\mathrm{CO}(J=1-0)$ PPV cubes.

\begin{figure}
    \centering
    \includegraphics[width=\columnwidth]{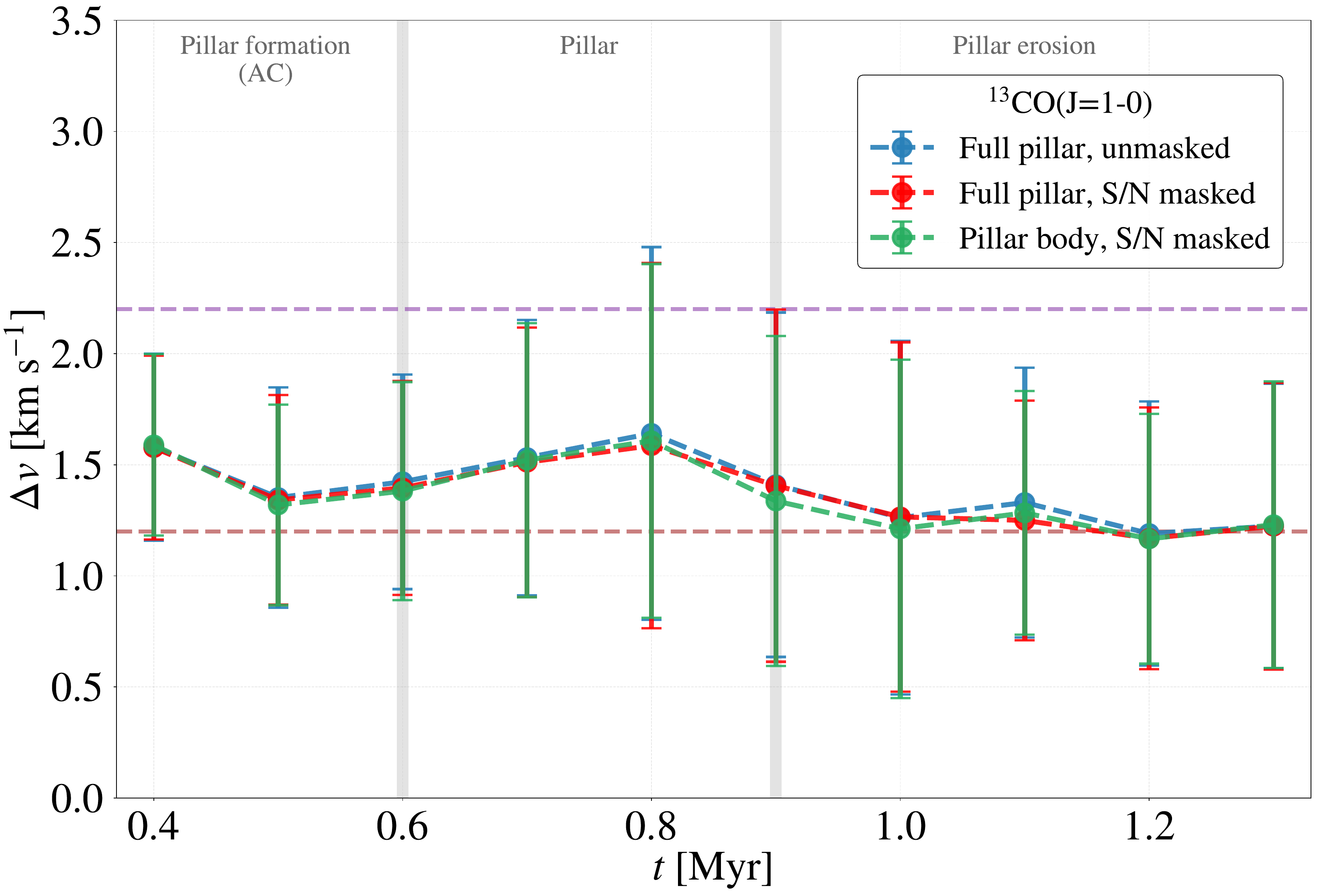} 
    \caption{Time evolution of the representative non-thermal line width ($\Delta v$) derived from the synthetic observations of the $^{13}\mathrm{CO}(J=1-0)$ molecular transition using the second spectral moment map (see Appendix~\ref{sec:appendix_secondOrderMomentmaps}).  The blue, red, and green dots show the values extracted for the three configurations used in the analysis: the full projected pillar without S/N masks, the full projected pillar with the polarimetric S/N masks ($p/\delta p > 3$ and $I/\delta I > 10$), and the S/N-selected pillar body after excluding the dynamically complex head region, respectively. The error bars correspond to the standard deviation of the spatial distribution of $\Delta v$ values derived from the second-moment maps within each selected region. The brown and magenta dashed horizontal lines represent the observational bounds of $\Delta v = 1.2$ and $2.2\,\mathrm{km\,s^{-1}}$ adopted for Pillar~II in M16 \citep[][ \citetalias{Pattle2018}]{White1999}. The vertical lines delimit the phases of pillar formation, evolution, and erosion, as in Fig.~\ref{fig:pillar_evol_ac_ratio}. The synthetic line widths obtained from all three configurations fall within the observational range reported for Pillar~II in M16.}
    \label{fig:linewidth_evolution}
\end{figure}

Fig.~\ref{fig:linewidth_evolution} shows the representative non-thermal line width $\Delta v$ throughout the different evolutionary stages of the simulated structure. The S/N-selected full pillar and the S/N-selected body-only measurements yield values that are not significantly different from the fiducial case.

During the stable lifetime of the pillar (from $t \approx 0.6$ to $0.9\,\mathrm{Myr}$), the spatially averaged non-thermal line widths in the fiducial configuration are measured to be $\Delta v \approx 1.42\pm0.48$, $ 1.53\pm0.62$, $1.64\pm0.84$, and $1.41\pm0.77\,\mathrm{km\,s^{-1}}$, respectively. The standard deviation of the 2D spatial distribution at each timestep is adopted as the uncertainty of the system, represented by the error bars in Fig.~\ref{fig:linewidth_evolution}.

These self-consistent kinematic measurements are consistent with the observational determinations. \citetalias{Pattle2018} adopted a line width range of $\Delta v \approx 1.2$ to $2.2\,\mathrm{km\,s^{-1}}$ for Pillar II in M16 (indicated by the horizontal dashed lines in Fig.~\ref{fig:linewidth_evolution}).

\subsection{Pillar mass and mean density}
\label{subsec:mass_density}

To self-consistently determine the mean density of the pillar, we first estimated its total mass from our synthetic molecular line observations. Using the \texttt{SOS} tool on the $^{13}\mathrm{CO}(J=1-0)$ PPV cubes from the synthetic observations generated with \texttt{LIME}, we evaluated the mass using the conversion factor (or X-factor) method ($M_{\mathrm{XF}}$) (Appendix~\ref{sec:appendix_mass_estimation}). To ensure a meaningful comparison, we computed the physical mass directly from the 3D R-MHD simulation grid ($M_{\mathrm{3D}}$) by integrating over the spatial mask that defines the physical boundaries of the pillar.

Figure~\ref{fig:mass_evolution} shows the time evolution of these mass estimates. During the stable lifetime of the pillar (from $t \approx 0.6$ to $0.9\,\mathrm{Myr}$), the mass measured directly from the simulation is $M_{\mathrm{3D}} \approx 107.91$, $99.29$, $96.59$, and $85.85\,\mathrm{M_{\odot}}$. The mass derived via the X-factor method yields $M_{\mathrm{XF}} \approx 152.71$, $120.80$, $96.71$, and $85.72 \,\mathrm{M_{\odot}}$, closely following the evolutionary trend of $M_{\mathrm{3D}}$.

Using $M_{\mathrm{XF}}$ as our observational proxy for the pillar mass, we computed the corresponding mean number density, $\langle n \rangle_{\mathrm{cyl}}$,\footnote{ { Note that $\langle n \rangle_{\mathrm{cyl}} = \langle \rho \rangle_{\mathrm{cyl}}/(\mu m_{\mathrm{H}})$, where $m_{\mathrm{H}}$ is the mass of the hydrogen atom. Since the simulations do not explicitly track the chemical evolution of the gas, we adopted a constant mean molecular weight of $\mu = 1.27$ \citep[see][]{ZA}.} } assuming the effective cylindrical geometry introduced in Section~\ref{sec:parameters}. The quoted uncertainties include the statistical errors propagated from the mass estimate.

\begin{figure}
    \centering
    \includegraphics[width=\columnwidth]{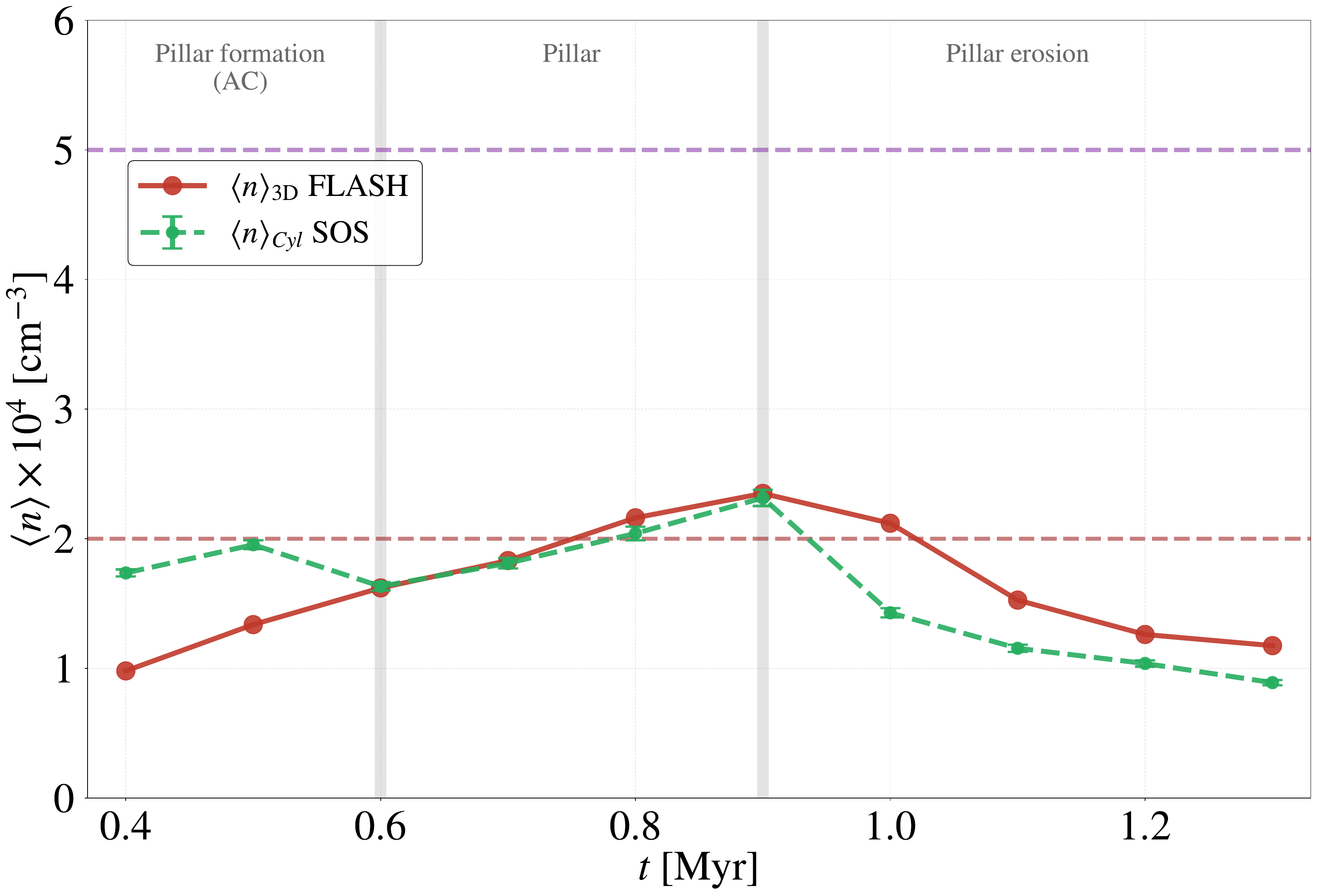} 
    \caption{Time evolution of the average number density of the pillar. The solid red line ($\langle n \rangle_{\mathrm{3D}}$) is measured from the simulation, while the green dashed line ($\langle n \rangle_{\mathrm{cyl}}$) is estimated self-consistently from the synthetic observations assuming an effective cylindrical geometry and the $M_{\mathrm{XF}}$ mass. The horizontal dashed lines represent the observational limits reported in the literature for Pillar II in M16 \citep[$\sim 2 \times 10^4$ to $5 \times 10^4\,\mathrm{cm^{-3}}$;][]{White1999, karim2023sofiaPillars}. The vertical lines delimit the phases of pillar formation, evolution, and erosion, as in Fig.~\ref{fig:pillar_evol_ac_ratio}. The mean density derived from the synthetic observations agrees well with the simulation during the stable pillar phase, although both remain a factor of $\sim$2--3 below the values reported for Pillar~II in M16.}
    \label{fig:mass_density_evolution}
\end{figure}

As shown in Fig.~\ref{fig:mass_density_evolution}, the observationally derived mean number density during the pillar phase ($\langle n \rangle_{\mathrm{cyl}} \approx 1.63$, $1.81$, $2.04$ and $2.32 \times 10^4\,\mathrm{cm^{-3}}$) is consistent with the 3D number density computed from the simulation ($\langle n \rangle_{\mathrm{3D}} \approx 1.62$, $1.83$, $2.16$, and $2.35 \times 10^4\,\mathrm{cm^{-3}}$). Furthermore, the $\langle n \rangle_{\mathrm{cyl}}$ values are in reasonable agreement with the observational constraints reported in the literature for Pillar II in M16, which range between $2 \times 10^4$ and $5 \times 10^4\,\mathrm{cm^{-3}}$ \citep[e.g.,][]{White1999, karim2023sofiaPillars}.

\begin{figure*}
\includegraphics[width=0.95\textwidth]{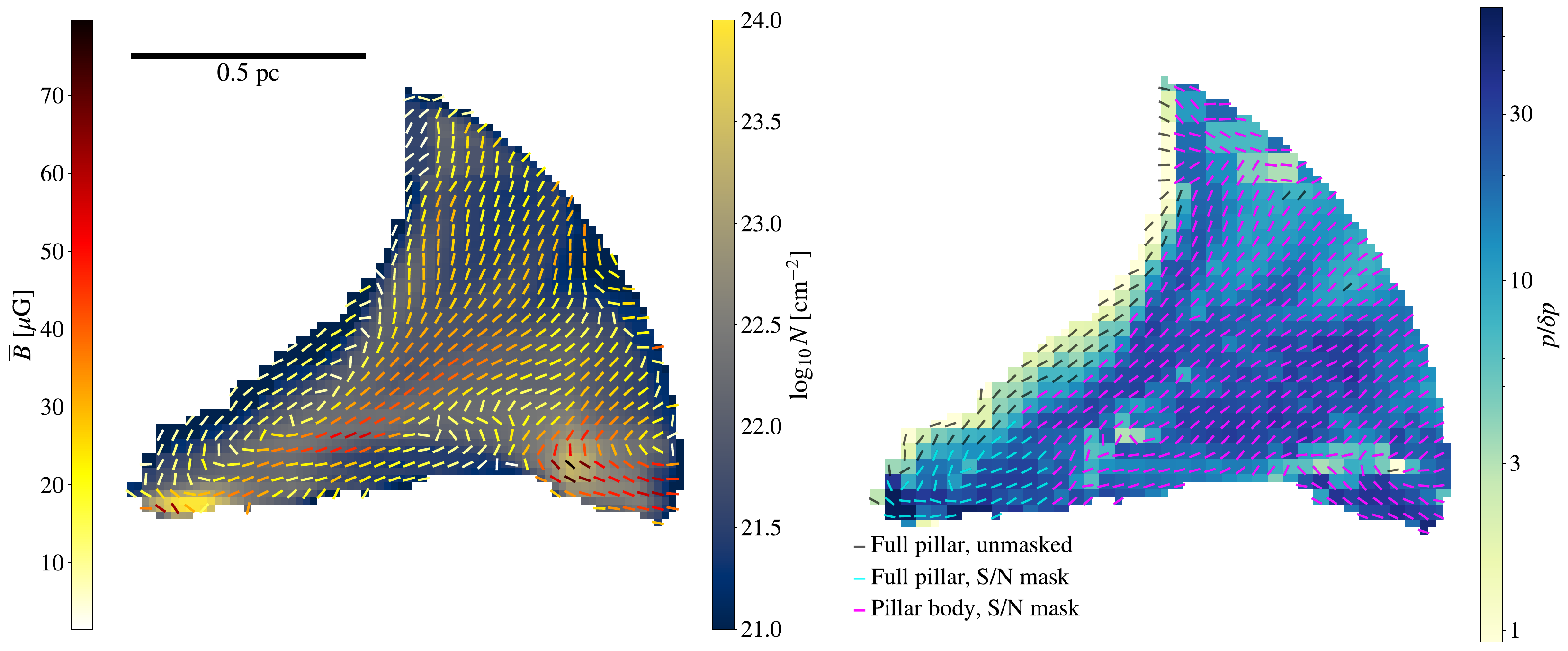}
\caption{Comparison between the intrinsic magnetic-field morphology of the simulation and the corresponding synthetic polarization observation at $t=0.7\,\mathrm{Myr}$. 
Left panel: column density map of the simulated pillar, shown as $\log_{10}N$ in units of $\mathrm{cm}^{-2}$, using the projected spatial footprint of the cold, dense, neutral gas. The pseudo-vectors show the density-weighted magnetic-field orientation projected onto the $z$--$y$ plane, while their colors indicate the line-of-sight density-weighted magnetic-field strength, $\overline{B}$.
Right panel: corresponding $p/\delta p$ map from the synthetic \texttt{POLARIS} observation. The color map is shown over the same projected spatial footprint used in the left panel and is not additionally masked by polarimetric S/N criteria. The pseudo-vectors are rotated by $90^{\circ}$ to indicate the inferred plane-of-sky magnetic-field direction. Black, cyan, and magenta pseudo-vectors indicate the three configurations used in the analysis: full pillar without S/N masking, full pillar with the polarimetric S/N mask ($I/\delta I>10$ and $p/\delta p>3$), and S/N-selected pillar body, respectively. The synthetic polarization morphology closely follows the intrinsic field structure in the simulation, while the simulated magnetic-field strength shows a clear gradient, becoming strongest along the dense spine and compressed lower/head regions of the pillar.}
\label{fig:pdpmapvsNdensmaps_142}
\end{figure*}

\subsection{Synthetic polarimetric observations of the pillar} \label{subsec:SPObs_pillar}

Using the classical DCF method, \citetalias{Pattle2018} reported plane-of-sky magnetic field strengths of $B_{\mathrm{POS}} = 170$--$320\,\mu\mathrm{G}$ for Pillar~II in M16. In contrast, from the raw simulation of our pillar (see Section~\ref{subsec:PillarIdentification}), we obtain significantly lower values, with an RMS (and mean) magnetic field strength of around $B_{\mathrm{rms, 3D}} \sim 33\,\mu\mathrm{G}$ (see Figure~\ref{fig:Bfield}). To investigate this discrepancy, we performed radiative transfer simulations of polarized dust emission at $850\,\mu\mathrm{m}$ using the \texttt{POLARIS} code, applied to the simulated pillar (see Section~\ref{subsec:so}) throughout its evolution.

From these synthetic observations, we obtained the Stokes parameters ($I$, $Q$, and $U$) to compute the polarization angle ($\psi$; Eq.~\ref{eq:polarisation_angle}) and the degree of linear polarization ($p$; Eq.~\ref{eq_linpolfrac}). Figure~\ref{fig:pdpmapvsNdensmaps_142} compares the magnetic field obtained directly from the simulation with the corresponding synthetic polarization map. The left panel shows the column density map derived from the simulation, overlaid with equal-length pseudo-vectors representing the density-weighted magnetic-field direction projected onto the $z$--$y$ plane.\footnote{Calculated along the line of sight ($x$-axis) as $\overline{B}_u = \sum_{i} B_{i,u} \rho_{i,x} / \sum_{i} \rho_{i,x}$ for $u \in \{x,y,z\}$, with a total magnitude $\overline{B} = (\overline{B}_{x}^2 + \overline{B}_{y}^2 + \overline{B}_{z}^2)^{1/2}$ and an orientation angle $\theta = \arctan(\overline{B}_{y} / \overline{B}_{z})$.} These pseudo-vectors are color-coded by the line-of-sight density-weighted magnetic-field strength, $\overline{B}$.

The right panel in Figure~\ref{fig:pdpmapvsNdensmaps_142} shows the S/N ratio map of the polarization fraction ($p/\delta p$), computed from the synthetic Stokes parameters and displayed over the same projected spatial footprint used for the simulation map. The color map is not restricted by the polarimetric S/N cuts, it shows the full projected pillar region used in our fiducial analysis. The overlaid pseudo-vectors show the inferred plane-of-sky magnetic-field direction, obtained by rotating the polarization vectors by $90^{\circ}$. Their colors indicate the three configurations used throughout the analysis: black for the full projected pillar without S/N masking, cyan for the full projected pillar with the polarimetric S/N mask ($I/\delta I>10$ and $p/\delta p>3$), and magenta for the S/N-selected pillar body after excluding the dynamically complex head region.

A visual comparison between the two panels suggests good qualitative agreement between the magnetic-field morphology traced by dust polarization and the intrinsic field obtained directly from the simulation. Overall, the synthetic polarization pseudo-vectors align with the magnetic-field lines along the major axis of the pillar, whereas a characteristic hairpin-like, or U-shaped, morphology is observed at the highly compressed tip. This magnetic-field alignment, driven by the expanding H\,{\sc ii} region, is consistent with observational measurements (e.g., \citetalias{Pattle2018}, \citealt{hwang2023}) and previous MHD simulations \citep[e.g.,][]{Arthur2011}.

To apply DCF-based methods self-consistently, we must estimate the characteristic polarization-angle dispersion ($\delta\psi$) entering the DCF relation. However, in environments shaped by external forces, large-scale morphological distortions (such as the U-shaped bending of the magnetic field around the pillar's head described above) artificially broaden the polarization-angle distribution. Low polarized-intensity regions can also increase the uncertainty of the inferred polarization angles, even in our idealized synthetic maps, which do not include instrumental noise. If not properly accounted for, this coherent field deformation may be misinterpreted as small-scale turbulent perturbations, thereby biasing the DCF estimates.

To assess the robustness of the inferred polarization-angle dispersion, we evaluated three spatial/observational configurations: the full projected pillar without S/N masking, the full projected pillar with the polarimetric S/N masks ($I/\delta I > 10$ and $p/\delta p > 3$), and the S/N-selected pillar body. We also compared two independent estimators: a Gaussian fitting analysis (G-Fit) and the Angular Dispersion Function (ADF; see Appendix~\ref{sec:appendix_adf_analysis} for details on the ADF implementation). Following \citet{Palau2021} and \citet{Polychronakis2025}, we avoid using the simple standard deviation of the polarization angles as our primary estimator because it can overestimate the turbulent dispersion when coherent large-scale field components are present. In our G-Fit analysis, the polarization-angle distributions are well described by a single Gaussian component for nearly all snapshots; therefore, we adopt the single-Gaussian width as our main estimate of $\delta\psi_{\mathrm{gfit}}$ (see Figure~\ref{fig:PAdist_gfits}).

\begin{figure}
    \centering
    \includegraphics[width=\columnwidth]{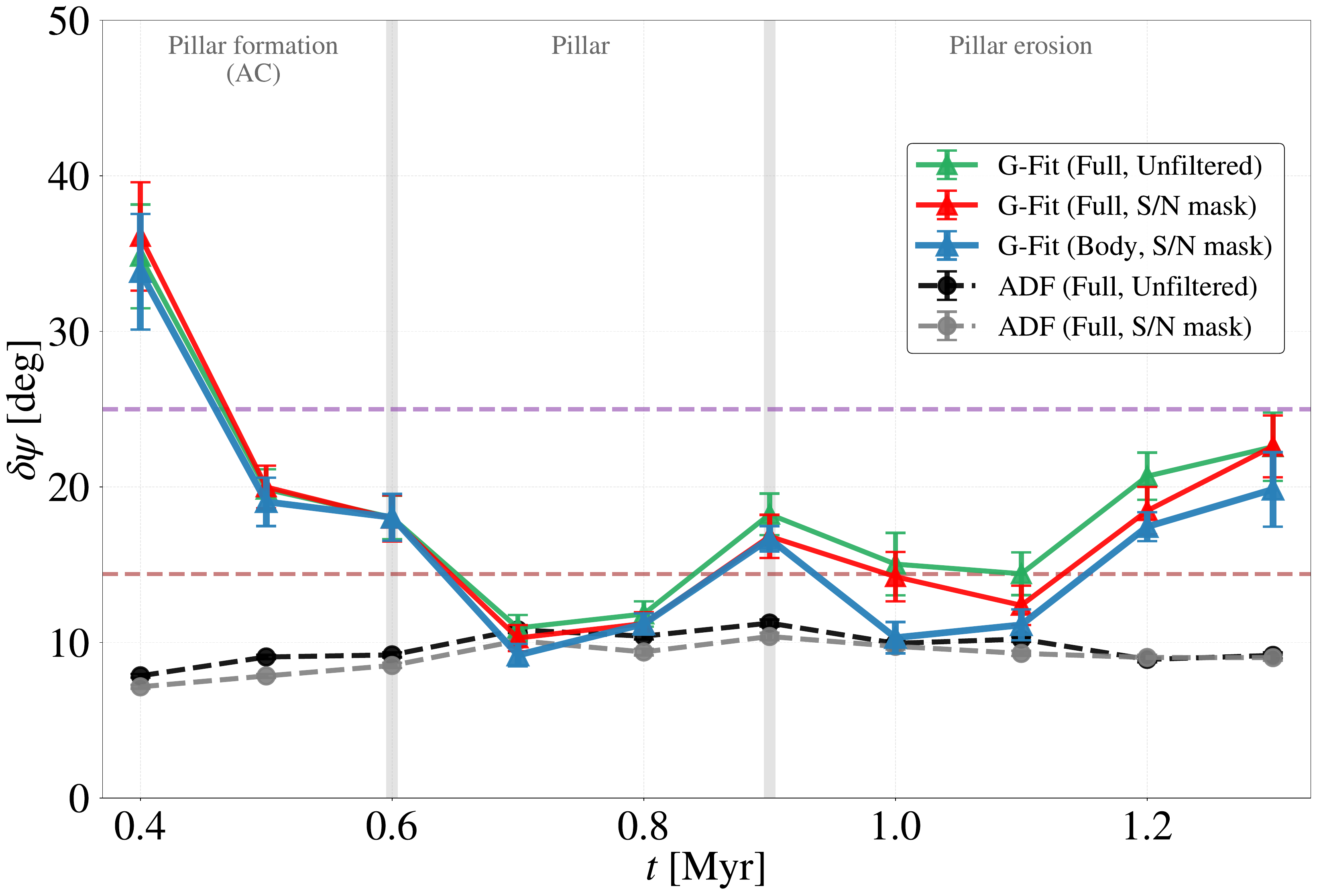} 
    \caption{ Time evolution of the polarization-angle dispersion, $\delta\psi$. Solid lines with triangular markers show the values obtained from the Gaussian fitting method, $\delta\psi_{\mathrm{gfit}}$, while dashed lines with circular markers show the ADF-derived dispersions. Green and black correspond to the full projected pillar without S/N masking; red and grey correspond to the full projected pillar with the polarimetric S/N mask ($I/\delta I>10$ and $p/\delta p>3$); and blue shows the G-Fit result for the S/N-selected pillar body. The full-pillar unmasked G-Fit measurement is adopted as the fiducial angular dispersion used in the DCF and ST21 magnetic-field estimates. The horizontal dashed lines indicate the value $\delta\psi=14.4^{\circ}$ adopted by \citetalias{Pattle2018} and the $25^{\circ}$ small-angle limit for the classical DCF method.}
    \label{fig:dispersion_evolution}
\end{figure}

Figure~\ref{fig:dispersion_evolution} shows the time evolution of the angular dispersion extracted under these different configurations. During the stable pillar phase (from $t = 0.6$ to $0.9$~Myr), our fiducial full-pillar unmasked measurements yield $\delta\psi_{\mathrm{gfit}} = 18.0\pm1.4^{\circ}$, $10.9\pm0.8^{\circ}$, $11.8\pm0.8^{\circ}$, and $18.2\pm1.3^{\circ}$, respectively. Applying the polarimetric S/N masks to the full projected pillar gives very similar values of $\delta\psi_{\mathrm{gfit}} = 18.0\pm1.5^{\circ}$, $10.3\pm0.8^{\circ}$, $11.2\pm0.7^{\circ}$, and $16.8\pm1.4^{\circ}$. Similarly, after removing the pillar head while retaining the S/N masks, the body-only measurements yield $\delta\psi_{\mathrm{gfit}} = 18.0\pm1.5^{\circ}$, $ 9.2\pm0.7^{\circ}$, $11.2\pm0.7^{\circ}$, and $16.7\pm0.8^{\circ}$.

These three G-Fit configurations therefore lead to comparable angular dispersions during the stable pillar phase. This indicates that neither the polarimetric S/N selection nor the exclusion of the pillar head significantly changes the inferred $\delta\psi_{\mathrm{gfit}}$. Consequently, we adopt the full projected pillar without S/N masking as our fiducial configuration for the self-consistent evaluation of the DCF-based methods.

As an independent methodological comparison, we also estimate $\delta\psi$ using the ADF method for the two full-pillar configurations shown in Fig.~\ref{fig:dispersion_evolution}. For the full unmasked pillar, this method gives $\delta\psi_{\mathrm{ADF}} = 9.2\pm0.2^{\circ}$, $10.8\pm0.2^{\circ}$, $10.4\pm0.2^{\circ}$, and $11.2\pm0.2^{\circ}$ from $t=0.6$ to $0.9$~Myr. For the full-pillar S/N-selected case, the corresponding values are $\delta\psi_{\mathrm{ADF}} =8.5\pm0.2^{\circ}$, $10.1\pm0.2^{\circ}$, $9.4\pm0.2^{\circ}$, and $10.4\pm0.2^{\circ}$. These ADF measurements are systematically lower than the Gaussian-fit estimates, with the difference largest near the phase boundaries ($t=0.6$ and $0.9\,\mathrm{Myr}$), where the single-Gaussian width is broadened by the onset and the disruption of the stable configuration; at the representative times $t=0.7$ and $0.8\,\mathrm{Myr}$ the two estimators agree to within $\sim1.5^{\circ}$. The ADF method is used only as an independent comparison of the angular-dispersion estimator and is not used to compute the DCF or ST21 magnetic-field strengths.

\subsection{Estimation of the plane-of-sky magnetic field} \label{subsec:bpos_estimation}

With the mean number density ($\langle n \rangle_{\mathrm{cyl}}$), the local non-thermal velocity dispersion ($\sigma_v$), and the characteristic polarization-angle dispersion ($\delta\psi_{\mathrm{gfit}}$) self-consistently derived from our synthetic observations, we can now evaluate the plane-of-sky magnetic field strength ($B_{\mathrm{POS}}$). We apply both the classical DCF method (Equation~\ref{eq_Bpos_clas_f05}) and the modified formulation by \citet{Skalidis2021} (ST21, Equation~\ref{eq:DCFmodifiedS21}). The corresponding uncertainties ($\Delta B_{\mathrm{POS}}$) are calculated through standard error propagation from the uncertainties in density, velocity dispersion, and the variance of the Gaussian fits for the angular dispersion.

\begin{figure*}
    \centering
    \includegraphics[width=0.95\textwidth]{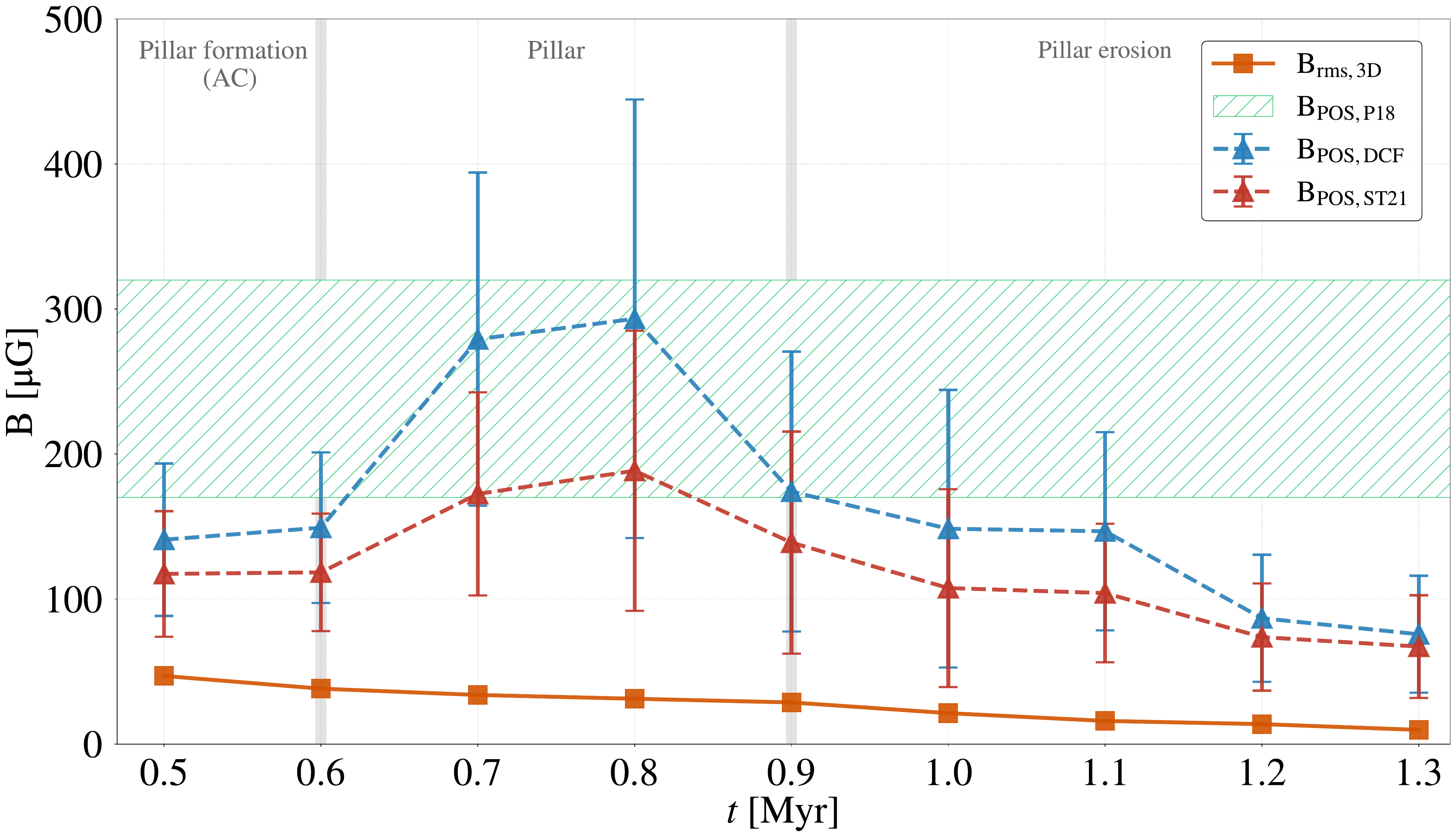} 
    \caption{Time evolution of the pillar plane-of-sky magnetic field strength ($B_{\mathrm{POS}}$). The triangles connected by dashed lines (dark blue for DCF and red for ST21) represent our self-consistent estimates ($B_{\mathrm{POS}}$). Error bars denote the uncertainties obtained via standard error propagation from the individual uncertainties in density, velocity dispersion, and the Gaussian variance of the angular dispersion (see Section~\ref{subsec:bpos_estimation}). The solid orange line indicates the RMS ($B_{\mathrm{rms, 3D}}$) intrinsic values extracted from the 3D simulation. The green horizontal band indicates the observational range reported by \citetalias{Pattle2018} for Pillar II in M16 ($170$--$320\,\mu\mathrm{G}$). The DCF-based estimates systematically overestimate the magnetic field strength relative to the intrinsic values measured directly from the simulation.}
    \label{fig:bpos_evolution}
\end{figure*}

Figure~\ref{fig:bpos_evolution} shows the time evolution of the derived plane-of-sky magnetic field strength using the classical DCF method ($B_{\mathrm{POS,DCF}}$, blue dashed line) and the modified ST21 method ($B_{\mathrm{POS,ST21}}$, red dashed line). For direct comparison, we also plot the magnetic field strengths extracted directly from the 3D R-MHD simulation grid, represented by the mean values ($B_{\mathrm{mean, 3D}}$, solid black line) and the root-mean-square values ($B_{\mathrm{rms, 3D}}$, solid orange line). In this figure, we show $B_{\mathrm{POS}}$ only for the time steps where the small-angle approximation is satisfied ($\delta\psi < 25^{\circ}$).

During the stable lifetime of the pillar, from $t \approx 0.6$ to $0.9\,\mathrm{Myr}$, our fiducial estimates using the classical DCF method yield $B_{\mathrm{POS,DCF}} \approx   149\pm52$, $279\pm115$, $293\pm151$, and $174\pm97\,\mu\mathrm{G}$. Comparing these values with the intrinsic field of the simulation, $B_{\mathrm{rms,3D}} \approx 38.2$, $33.8$, $31.2$, and $28.7\,\mu\mathrm{G}$, we find that the DCF estimates overestimate $B_{\mathrm{rms,3D}}$ by factors of $\sim 3.9$, $\sim 8.3$, $\sim 9.4$, and $\sim 6.1$, respectively, with an average factor of $\sim 7$.

A similar, although somewhat smaller, discrepancy is obtained when the incompressibility assumption is relaxed. For the ST21 method, the fiducial full-pillar measurements give $B_{\mathrm{POS,ST21}} \approx  118\pm40$, $172\pm70$, $188\pm97$, and $139\pm77\,\mu\mathrm{G}$. These values exceed $B_{\mathrm{rms,3D}}$ by factors of $ \sim 3.1$, $\sim 5.1$, $\sim 6.0$, and $\sim 4.8$, respectively, corresponding to an average factor of $\sim 5$.

To quantify the sensitivity of these results to the adopted observational selection, we repeated the calculation for two additional configurations: the full projected pillar with the polarimetric S/N masks, and the S/N-selected considering only the pillar body. These tests are summarized at the representative time $t=0.7\,\mathrm{Myr}$ in Table~\ref{tab:dcf_decomposition}. The three configurations give comparable angular dispersions, line widths, and inferred field strengths. In particular, at $t=0.7\,\mathrm{Myr}$, the DCF estimates are $279\pm115$, $ 293\pm120$, and $ 330\pm136\,\mu\mathrm{G}$ for the full unmasked, full S/N-selected, and body S/N-selected cases, respectively. The corresponding ST21 estimates are $172\pm70$, $175\pm71$, and $187\pm76\,\mu\mathrm{G}$. Across all three configurations, the inferred field strengths remain systematically above the intrinsic simulation value, indicating that the DCF/ST21 overestimation is not removed by either the application of polarimetric S/N masks or the adopted spatial selection (full pillar versus pillar body).

\begin{table*}
    \centering
    \caption{Decomposition of the DCF and ST21 overestimation at the representative time $t = 0.7\,\mathrm{Myr}$. The fiducial configuration is the full projected pillar without applying polarimetric S/N masks. The two additional configurations test the sensitivity of the results to S/N selection and to the spatial exclusion of the dynamically complex pillar head. All configurations use the same mean gas density, while the line width and polarization-angle dispersion are measured for each analyzed region. The raw simulation and the observed values for Pillar~II in M16 are listed for reference.}
    \label{tab:dcf_decomposition}
    \setlength{\tabcolsep}{5pt}
    {
    \begin{tabular}{l c c c c c c}
        \hline
        Configuration & $\Delta v$ (km s$^{-1}$) & $\delta\psi_{\mathrm{gfit}}$ ($^{\circ}$) & $B_{\mathrm{POS,DCF}}$ ($\mu$G) & $B_{\mathrm{POS,DCF}}/B_{\mathrm{rms,3D}}$ & $B_{\mathrm{POS,ST21}}$ ($\mu$G) & $B_{\mathrm{POS,ST21}}/B_{\mathrm{rms,3D}}$ \\
        \hline
        Full pillar, unmasked (fiducial) & $1.53\pm0.62$ & $10.9\pm0.8$ & $279\pm115$ & $8.3$ & $172\pm70$ & $5.1$ \\
        Full pillar, S/N mask & $1.51\pm0.61$ & $10.3\pm0.8$ & $293\pm120$ & $8.7$ & $175\pm71$ & $5.2$ \\
        Pillar body, S/N mask & $1.52\pm0.62$ & $9.2\pm0.7$ & $330\pm136$ & $9.8$ & $187\pm76$ & $5.5$ \\
        Raw simulation ($B_{\mathrm{rms,3D}}$) & $\cdots$ & $\cdots$ & $33.8$ & $1$ & $33.8$ & $1$ \\
        Observed (P18, Pillar II)$^{a}$ & $1.2$--$2.2$ & $14.4$ & $170$--$320$ & $\cdots$ & $\cdots$ & $\cdots$ \\
        \hline
    
    \end{tabular}
    }
    \begin{flushleft}
        \footnotesize{  All quantities are evaluated at $t = 0.7\,\mathrm{Myr}$; their full temporal evolution is shown in Figs.~\ref{fig:dispersion_evolution},~\ref{fig:linewidth_evolution}, and~\ref{fig:bpos_evolution}.\\
        $^{a}$ Observed values for Pillar~II in M16 \citepalias{Pattle2018}.}
    \end{flushleft}
    
\end{table*}

This systematic discrepancy therefore indicates that the assumptions underlying DCF-based methods break down in pillar-like structures shaped by external compression. The DCF method assumes a relation between internal turbulent kinetic fluctuations and turbulence-driven distortions of the magnetic field. However, in externally compressed pillars, the expanding \HII\ region imposes a coherent, pillar-scale magnetic morphology through transverse and axial compression. This compression both amplifies the ordered field component and suppresses small-scale angular fluctuations. Simultaneously, the externally driven compression increases the dispersion of the velocity field, decoupling the measured gas motions from the polarization-angle perturbations that enter the DCF estimate. As a result, the measured polarization-angle dispersion no longer provides a direct measure of turbulence-driven magnetic distortions, leading DCF-based methods to systematically overestimate the inferred magnetic field strength.

\section{Discussion}\label{sec:discussion}

In this study, we used synthetic observations to assess the reliability of $850\,\mu\mathrm{m}$ dust polarization as a tracer of magnetic fields in pillars around \HII\ regions, and to evaluate the accuracy of the Davis-Chandrasekhar-Fermi (DCF) method in these environments shaped by external compressions.

By comparing the 3D morphology obtained from our simulation with synthetic observations, we find that the linear polarization of thermal dust emission provides a reliable tracer of the underlying magnetic field geometry. In general, the polarization { pseudo--}vectors align with the magnetic field lines along the major axis of the pillar, while a characteristic hairpin-shaped morphology is observed at the highly compressed tip, consistent with other numerical simulations of R-MHD \citep[e.g.,][]{Arthur2011}. Furthermore, our parametric study using the \texttt{POLARIS} radiative transfer code \citep{Reissl2016} shows that the orientation of the synthetic polarization vectors is mathematically invariant to variations in the dust alignment efficiency, such as the fraction of grains of high state $J$ ($f_{\mathrm{high-}J}$) or the inclusion of internal thermal fluctuations (see Appendix~\ref{sec:appendix_polaris_convergence}). Because the Rayleigh reduction factor scales both Stokes $Q$ and $U$ equally, its dependence cancels out completely when calculating the polarization angle $\psi$. Convergence tests with high ray-tracing resolutions confirm that the observed highly ordered magnetic field morphology is an intrinsic physical feature of the R-MHD simulated pillar, effectively ruling out spatial subsampling artifacts.

It is important to note that our synthetic observations represent an idealized ``best-case'' scenario, as they do not include instrumental noise or beam-smoothing effects. Early numerical tests by \citet{heitsch2001_DCFtestsInNumSims} and more recent synthetic observations \citep[e.g.,][]{Juarez+2017} have shown that limited telescope resolution and spatial subsampling can smooth out small-scale fluctuations, thereby biasing the estimated field strength while largely preserving the global polarization morphology. At the same time, real polarization maps can show additional apparent dispersion due to noise, line-of-sight contamination, and incomplete spatial sampling. Consequently, a visual comparison between observed polarization maps and our synthetic observations may suggest differences that reflect observational limitations rather than a physical disagreement. In their analysis of Pillar II in M16, \citetalias{Pattle2018} selected the highest S/N pixels ($I/\delta I > 10$ and $p/\delta p > 3$) and calculated an angular dispersion of $\delta\psi = 14.4^{\circ}$.

We therefore stress that the similarity between the masked and unmasked synthetic measurements should not be interpreted as evidence that polarimetric S/N cuts are generally unimportant. Previous observational analyses have shown that poor sensitivity and sparse sampling of the polarized emission can introduce large uncertainties in the polarization-angle dispersion and, consequently, in DCF-based magnetic-field estimates \citep[e.g.,][]{liu2019ApJ, Palau2021}. Our result only shows that, within this controlled synthetic setup and for the adopted masks, the DCF/ST21 overestimation is not removed by applying the polarimetric S/N selection.

Despite qualitative morphological agreement, applying the DCF method to such structures presents challenges, since morphological agreement does not guarantee the validity of the underlying quantitative assumptions. The classical DCF method \citep{davis1951, chandrasekhar1953} assumes that angular dispersion arises solely from fluctuations about a large-scale uniform background field driven by isotropic turbulence. Importantly, in their analysis, \citetalias{Pattle2018} implicitly assumed that all measured angular dispersion represents fluctuations about a largely uniform mean field direction. However, in environments shaped by external forces, large-scale morphological distortions (such as the strong U-shaped magnetic field bending around the pillar's head) contribute additional angular spread to the global distribution that is not turbulent in origin. If this distinction is not considered, part of the macroscopic field morphology can be interpreted as a turbulent perturbation.

To evaluate this effect self-consistently, we measured the polarization-angle dispersion using Gaussian fits to the angular distributions, rather than adopting the simple standard deviation of the polarization angles. Following the motivation of \citet{Palau2021} and \citet{Polychronakis2025}, this choice avoids assigning excessive weight to extended tails or coherent large-scale components when estimating the characteristic dispersion. In practice, nearly all of our fitted distributions are best described by a single Gaussian component, so we adopt the single-Gaussian width as our main estimate of $\delta\psi_{\mathrm{gfit}}$.

For the quantitative analysis, we adopt the full projected pillar without applying polarimetric S/N masks as our fiducial configuration. During the stable pillar phase, from $t \approx 0.6$ to $0.9\,\mathrm{Myr}$, the fiducial angular dispersions are $\delta\psi_{\mathrm{gfit}} \approx 18.0\pm1.4^{\circ}$, $ 10.9\pm0.8^{\circ}$, $11.8\pm0.8^{\circ}$, and $ 18.2\pm1.3^{\circ}$. These values are comparable to the dispersion of $14.4^{\circ}$ reported by \citetalias{Pattle2018} for Pillar II in M16 and remain within the small-angle regime required for the DCF analysis. To ensure a self-consistent evaluation of the magnetic field strength, we complement this angular dispersion with the mean number density ($\langle n \rangle_{\mathrm{cyl}}$) and the non-thermal velocity dispersion ($\sigma_v$), both extracted from $^{13}\mathrm{CO}$ synthetic molecular line observations as described in Sections~\ref{subsec:velocity_dispersion} and~\ref{subsec:mass_density}.

Our results show that even when the three DCF input parameters are self-consistently and locally extracted, the analytical methods overestimate the intrinsic plane-of-sky magnetic field strength by large factors. As shown in Section~\ref{subsec:bpos_estimation}, the classical DCF approach gives fiducial values of $B_{\mathrm{POS,DCF}} \approx  149\pm52$, $ 279\pm115$, $ 293\pm151$, and $174\pm97\,\mu\mathrm{G}$ during the stable pillar phase. Compared with the directly simulated values $B_{\mathrm{rms,3D}} \approx 38.2$, $33.8$, $31.2$, and $28.7\,\mu\mathrm{G}$, these estimates correspond to overestimation factors of $ \sim 3.9$, $ \sim 8.3$, $ \sim 9.4$, and $ \sim 6.1$, respectively, with an average factor of $\sim 7$.

The two additional observational selections give similar results. At the representative time $t=0.7\,\mathrm{Myr}$, the full unmasked, full S/N-selected, and body S/N-selected configurations yield $B_{\mathrm{POS,DCF}} \approx   279\pm115$, $ 293\pm120$, and $ 330\pm136\,\mu\mathrm{G}$, respectively. The corresponding ST21 estimates are $B_{\mathrm{POS,ST21}} \approx  172\pm70$, $  175\pm71$, and $ 187\pm76\,\mu\mathrm{G}$. Thus, the inferred overestimation is not primarily driven by the polarimetric S/N selection or by the removal of the pillar head. Instead, it is already present in the full projected pillar, which we adopt as the fiducial configuration.

Ultimately, this systematic overestimation indicates that DCF-based estimates should be interpreted with caution in externally compressed pillar-like structures. Interestingly, \citet{liu2021calibrating} showed that the DCF method performs well in cases of strong magnetic fields but can significantly overestimate field strengths in weaker field regimes where energy equipartition is not reached. While our pillar exhibits moderate intrinsic field strengths ($B_{\mathrm{rms,3D}} \lesssim 40\,\mu\mathrm{G}$ during its stable phase), the fundamental reason for the DCF overestimation lies in its externally driven nature. The large-scale magnetic field alignment is driven primarily by the external ram and thermal pressures of the expanding H\,{\sc ii} region. Because these external agents continually govern the field morphology throughout the pillar's evolution, the system inherently lacks the isolated internal energy equipartition assumed by the DCF method.

In this context, the measured velocity and polarization-angle dispersions do not trace the same statistically coupled perturbations assumed by DCF-based methods. The expansion of the H\,{\sc ii} region produces coherent compression and stretching motions that contribute to the measured gas velocity dispersion, while organizing the magnetic field over pillar scales. Consequently, $\sigma_v$ contains a substantial contribution from non-turbulent motions, whereas $\delta\psi$ does not provide a corresponding measure of the magnetic perturbations associated with those motions. This decoupling increases the ratio $\sigma_v/\delta\psi$ entering the classical DCF estimate and leads to a systematic overestimate of $B_{\mathrm{POS}}$. The ST21 formulation reduces the sensitivity to small angular dispersions through its $\delta\psi^{-1/2}$ dependence, but still requires the measured velocity and angular dispersions to represent physically related perturbations. This condition is not fully satisfied in the externally compressed pillar.

We emphasize that this work constitutes a controlled case study of a self-consistently formed pillar followed throughout its evolution in an R-MHD simulation. The synthetic maps are designed to isolate the underlying physical effects and therefore do not include an explicit realization of instrumental noise or beam convolution; moreover, they are analysed along a single line of sight. Within this framework, our results demonstrate a persistent systematic overestimation of the magnetic-field strength, although the precise factors reported here may vary with pillar geometry, evolutionary stage, viewing angle, numerical resolution, and observational setup. Extending this analysis to a statistical sample spanning a broader range of pillars and physical conditions will be necessary to determine the domain over which these quantitative factors apply.

Consequently, DCF-derived magnetic field strengths around H\,{\sc ii} regions (such as pillars and similar externally compressed dense structures) should be treated with caution, as the present results show that they can represent significant overestimations of the magnetic field's true strength and its derived dynamical quantities (e.g., magnetic pressure, Alfvén Mach number, etc). These results provide a complementary physical explanation for the systematic tendency reported in the literature that DCF-derived strengths are typically a factor of $\sim 3$--$5$ higher than those derived from Zeeman effect measurements \citep[see e.g., Section 6.1.1 of][]{pattle2022magnetic}; though we note that this discrepancy likely has multiple contributing causes beyond the specific mechanism identified here. Alternative methods \citep[e.g.,][]{LQZ22} or dedicated analytical corrections \citep[e.g.,][]{liu2021calibrating, Chen2022} that explicitly account for anisotropic compression and non-turbulent field alignment are strongly recommended to properly assess the role of magnetic fields in the evolution of pillar-like and other externally compressed structures.

\section{Summary and Conclusions}\label{sec:conclusions}

In this work, we used three-dimensional radiation-magnetohydrodynamic (R-MHD) simulations combined with synthetic polarimetric and molecular line observations (\texttt{POLARIS} and \texttt{LIME} codes) to evaluate the reliability of dust polarization at $850\,\mu\mathrm{m}$ as a magnetic field tracer and to test the accuracy of the Davis--Chandrasekhar--Fermi (DCF) method in pillar-like structures at the boundaries of H\,{\sc ii} regions, where the field geometry is driven primarily by external compression rather than internal turbulence.

We focus on the first pillar that forms self-consistently in our simulation as a proof of concept. The expanding ionization front compresses a dense molecular clump embedded in a molecular filament, producing a pillar-like morphology through a combination of transverse and axial compressions. During the pillar phase, the magnetic field becomes aligned with the main axis of the structure, providing partial support against further collapse and maintaining the overall shape as the external pressure weakens  \citep[see][]{ZamoraAvils2019}.

To estimate the magnetic field strength using DCF-type methods, we self-consistently derived the relevant pillar properties throughout its evolution, namely the mean density $\langle n \rangle_{\mathrm{cyl}}$, the non-thermal velocity dispersion $\sigma_v$, and the polarization-angle dispersion $\delta\psi_{\mathrm{gfit}}$. We adopt the full projected pillar without applying polarimetric S/N masks as our fiducial configuration. Our main findings are summarized as follows:

\begin{enumerate}
    \item{Synthetic observations of polarized dust emission at $850\,\mu$m successfully reproduce the large-scale magnetic morphology of the pillar. A comprehensive parametric study (Appendix~\ref{sec:appendix_polaris_convergence}) shows that the orientation of the synthetic polarization vectors is mathematically invariant to the assumed dust alignment efficiency. This confirms that the highly ordered magnetic field morphology is an intrinsic physical feature of the R-MHD simulation and that dust polarimetry is a reliable diagnostic of the magnetic field geometry in these environments.}

    \item{The polarization-angle distributions are, in practice, well described by a single Gaussian component for nearly all snapshots. We therefore adopt the single-Gaussian width as our main estimate of $\delta\psi_{\mathrm{gfit}}$. For the fiducial configuration, the stable pillar phase ($t \approx 0.6$--$0.9\,\mathrm{Myr}$) yields $\delta\psi_{\mathrm{gfit}} \approx 18.0\pm1.4^{\circ}$, $10.9\pm0.8^{\circ}$, $11.8\pm0.8^{\circ}$, and $18.2\pm1.3^{\circ}$. The S/N-selected full-pillar and body-only measurements give comparable values, indicating that the inferred angular dispersion is not strongly controlled by the adopted observational mask or by the exclusion of the pillar head.}  

    \item{ Even when all DCF input parameters are extracted self-consistently and locally from the simulation, the inferred magnetic field strengths differ substantially from the intrinsic simulation values. For the fiducial configuration, the classical DCF method \citep{davis1951, chandrasekhar1953} overestimates the true field by an average factor of $  {\sim 7}$ during the stable pillar phase, while the modified formulation of \citet{Skalidis2021} reduces the discrepancy but still overestimates the field by an average factor of $  {\sim 5}$. At the representative time $t=0.7\,\mathrm{Myr}$, the full unmasked, full S/N-selected, and body S/N-selected configurations yield comparable values: $B_{\mathrm{POS,DCF}} \approx   279\pm115$, $ 293\pm120$, and $  330\pm136\,\mu\mathrm{G}$, respectively, while the corresponding ST21 estimates are $B_{\mathrm{POS,ST21}} \approx 172\pm70$, $ 175\pm71$, and $187\pm76\,\mu\mathrm{G}$. This demonstrates that the overestimation is not an artefact of the particular spatial or S/N selection adopted for the analysis.}

    \item{ The origin of this overestimation is physical rather than methodological. The magnetic-field geometry and amplification in the simulated pillar are governed primarily by the external thermal and ram pressure of the expanding H\,{\sc ii} region, not by internal turbulence alone. This violates the core DCF assumption that the observed angular dispersion traces turbulent fluctuations about an internally generated, approximately uniform background field. Within the pillar, externally driven compression and the associated coherent motions contribute to the measured gas velocity dispersion, while simultaneously aligning and amplifying the magnetic field on pillar scales. The resulting ordered field morphology maintains a relatively small $\delta\psi$; consequently, $\sigma_v$ and $\delta\psi$ do not trace the same statistically coupled perturbations assumed by the DCF method. This mismatch biases the ratio $\sigma_v/\delta\psi$ high, leading to a systematic overestimate of $B_{\mathrm{POS}}$.}
\end{enumerate}

In conclusion, the systematic overestimation of the magnetic field strength shown here arises because the DCF method assumes that the angular dispersion traces internal turbulent fluctuations of a uniform background field, a condition that is not met when the field geometry is governed by external compression. As a result, DCF-based field strengths in these environments, and potentially in other externally driven structures, should be interpreted with caution, since they may translate into overestimated dynamical quantities such as the magnetic pressure or the Alfvén Mach number. We stress, however, that this assessment is based on a single simulated pillar and the broader study outlined below is required before the reported factors can be generalized.

Future work will extend this analysis to a broader set of simulated pillars that span a range of physical conditions to assess how the overestimation factor depends on them. Exploring these regimes will allow for a broader evaluation of the DCF method's validity and a better understanding of the complex interplay between magnetic fields, radiative feedback, and cloud structure in star-forming environments.

\section*{Acknowledgements}
The authors thank the anonymous referee, whose constructive suggestions helped strengthen the manuscript and clarify its main results. 
A.P. acknowledges financial support from the UNAM-PAPIIT IN120226 grant. AP and MZA also acknowledge support from the SNII of SECIHTI, M\'exico. The authors thankfully acknowledge the computer resources, technical advice, and support provided by LANCAD-UNAM.DGTIC-188 and SECIHTI, through the use of the Miztli supercomputer at DGTIC–UNAM.

\section*{Data Availability}

The data generated for this article will be shared on request to the corresponding author.



\bibliographystyle{mnras}
\bibliography{refs} 



%
\newpage
\appendix

\section{Total mass estimation}
\label{sec:appendix_mass_estimation}

\begin{figure}
    \centering
    \includegraphics[width=\columnwidth]{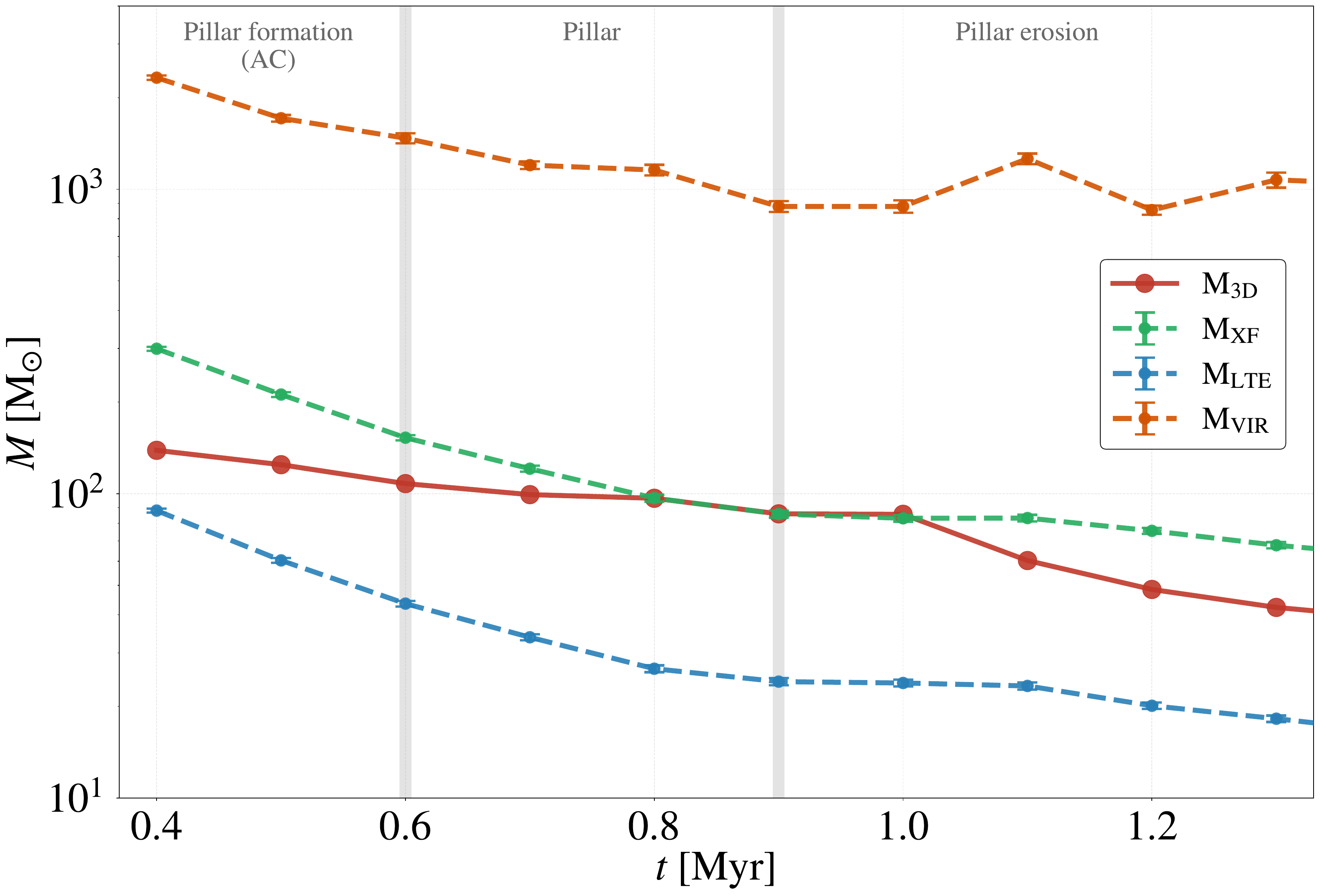} 
    \caption{Time evolution of the total mass of the pillar. The solid red line ($M_{\mathrm{3D}}$) represents the mass computed directly from the 3D R-MHD simulation. The dashed lines show the mass estimates derived from the synthetic $^{13}\mathrm{CO}(J=1-0)$ observations calculated using the \texttt{SOS} tool under three different approximations: the conversion factor (X-factor) method ($M_{\mathrm{XF}}$, green line), the assumption of local thermodynamic equilibrium ($M_{\mathrm{LTE}}$, blue line), and the virial theorem ($M_{\mathrm{vir}}$, orange line). The vertical lines delimit the phases of pillar formation, evolution, and erosion, as in Fig.~\ref{fig:pillar_evol_ac_ratio}. The $M_{\mathrm{XF}}$ estimate follows the simulation mass most closely.}
    \label{fig:mass_evolution}
\end{figure}

We processed the synthetic molecular line observations using the \texttt{SOS} tool to evaluate the pillar's physical properties via three independent mass estimation methods: Local Thermodynamic Equilibrium (LTE, $M_{\mathrm{LTE}}$), the virial theorem ($M_{\mathrm{vir}}$), and the X-factor method ($M_{\mathrm{XF}}$) \citep[see e.g.,][]{RetesRomero2017}.

For all estimates, the effective radius ($R_e$) is evaluated geometrically, assuming a distance $d = 1000\,\mathrm{pc}$ (consistent with the setup adopted in the \texttt{POLARIS} simulations), an angular diameter $\theta \approx 0.229^{\circ}$, and a scaling factor $k_{R_e} = 1$:
\begin{equation}
    R_e = k_{R_e} \cdot d \cdot \tan\left(\frac{\theta}{2} \frac{\pi}{180^{\circ}}\right).
\end{equation}
We assumed null geometric uncertainty ($\Delta R_e = 0$) and a $25\%$ systematic uncertainty for abundance and conversion factors ($\Delta X / X = 0.25$).

\subsection{Local Thermodynamic Equilibrium mass (\texorpdfstring{$M_{\mathrm{LTE}}$}{M\_LTE})}

The LTE mass integrates the column density of the optically thin tracer ($N_{\mathrm{thin}}$ for $^{13}\mathrm{CO}$) scaled by its relative abundance:
\begin{equation}
    M_{\mathrm{LTE}} = 3.25 \cdot \left(\frac{R_e}{5.0\,\mathrm{pc}}\right)^2 \cdot \left(\frac{N_{\mathrm{thin}}}{10^{17}\,\mathrm{cm}^{-2}}\right).
\end{equation}
Assuming the thermal noise ($\Delta T_{\mathrm{thin}} = \sigma_{\mathrm{rms}}$) dominates the local observational uncertainty, the statistical error propagates as:
\begin{equation}
    \Delta M_{\mathrm{LTE}} = M_{\mathrm{LTE}} \sqrt{ \left(\frac{\Delta X}{X}\right)^2 + \left(\frac{\Delta T_{\mathrm{thin}}}{T_{\mathrm{thin}}}\right)^2 }.
\end{equation}

\subsection{Virial mass (\texorpdfstring{$M_{\mathrm{vir}}$}{M\_vir})}

The virial mass assumes perfect gravitational equilibrium supported solely by the internal gas velocity dispersion, parameterized by the full width at half maximum of the $^{13}\mathrm{CO}$ line profile ($\mathrm{FWHM} = \Delta V$) \citep{RetesRomero2017}:
\begin{equation}
    M_{\mathrm{vir}} = 1.58 \cdot \left(\frac{R_e}{5.0\,\mathrm{pc}}\right) \cdot \left(\frac{\Delta V}{5.0\,\mathrm{km\,s}^{-1}}\right)^2.
\end{equation}
The corresponding statistical uncertainty is:
\begin{equation}
    \Delta M_{\mathrm{vir}} = M_{\mathrm{vir}} \sqrt{ \left(\frac{\Delta X}{X}\right)^2 + \left(2 \frac{\Delta (\Delta V)}{\Delta V}\right)^2 }.
\end{equation}

\subsection{X-Factor mass (\texorpdfstring{$M_{\mathrm{XF}}$}{M\_XF})}

The X-factor mass is based on the integrated intensity of the tracer's line profile ($W_{^{13}\mathrm{CO}} = \int T_{^{13}\mathrm{CO}} dv$), where $T_{^{13}\mathrm{CO}}$ is the brightness temperature of the emission, and $dv$ is the velocity differential (which corresponds to the channel width in our PPV cubes). We customized the \texttt{SOS} mass expression for our specific environment:
\begin{equation}
    M_{\mathrm{XF}} = \pi \mu_2 m_{\mathrm{H}} \mathrm{[^{12}CO/^{13}CO]} X_{\mathrm{F}} W_{^{13}\mathrm{CO}} R_e^2,
\end{equation}
where $\mu_2 = 1.27$ is the mean molecular weight, $m_{\mathrm{H}}$ is the mass of hydrogen atoms, and $\mathrm{[^{12}CO/^{13}CO]} = 44.65$ is the isotopic abundance ratio \citep{karim2023sofiaPillars}. Although the large-scale M16 cloud is characterized by an average conversion factor of $X_{\mathrm{F}} = 2.15 \times 10^{20}\,\mathrm{cm^{-2}\,K^{-1}\,km^{-1}\,s}$ \citep{kohno2024co}, that value is representative of the cloud as a whole and is not necessarily appropriate for the local physical conditions of the pillar analyzed here. We therefore adopt a local value of $1.0 \times 10^{20}\,\mathrm{cm^{-2}\,K^{-1}\,km^{-1}\,s}$, which remains within the range of observationally plausible values for dense molecular substructures and provides a better match to the independently measured pillar mass during its stable lifetime ($t \approx 0.6$--$0.9,\mathrm{Myr}$; see the dashed green line in Figure~\ref{fig:mass_evolution}).

CO luminosity is known to depend on the gas velocity dispersion and temperature \citep{narayanan2012, bolatto2013co}. In our case, the external thermal and ram pressures exerted by the expanding H\,{\sc ii} region are expected to heat the pillar boundaries and enhance the local gas motions. Under these conditions, CO emission can be brighter per unit of H$_2$ mass than in the more quiescent cloud-averaged environment, making a lower local $X_{\mathrm{F}}$ physically plausible for the pillar. Consequently, we estimate the molecular hydrogen column density as

\begin{equation}
N(\mathrm{H}_2) = \mathrm{[^{12}CO/^{13}CO]} \cdot X_{\mathrm{F}} \cdot W_{^{13}\mathrm{CO}}.
\end{equation}

The uncertainty of the integrated flux ($\Delta W$) propagates from the RMS noise per-pixel ($\sigma_{\mathrm{rms}}$), spectral resolution ($\delta v$), and number of integrated channels ($N_{\mathrm{chns}}$):
\begin{equation}
    \Delta W = \sigma_{\mathrm{rms}} \cdot \delta v \cdot \sqrt{N_{\mathrm{chns}}}.
\end{equation}
{ The local statistical error for $M_{\mathrm{XF}}$ then becomes:}
\begin{equation}
    \Delta M_{\mathrm{XF}} = M_{\mathrm{XF}} \sqrt{ \left(\frac{\Delta X}{X}\right)^2 + \left(\frac{\Delta W}{W}\right)^2 }.
\end{equation}

For all methods, the total mass was estimated by discretizing the projected structure into a $64 \times 64$ spatial grid. Within each cell, \texttt{SOS} fits a local Gaussian to the spectrum in order to derive the linewidth, antenna temperature, and integrated flux, from which the corresponding local estimates of $M_{\mathrm{LTE}}$, $M_{\mathrm{vir}}$, and $M_{\mathrm{XF}}$ are obtained. The total mass of the pillar is then computed as the sum over all valid cells within the spatial mask, $M_{\mathrm{total}} = \sum M_{\mathrm{bin}}$, while the final uncertainty is calculated by propagating the individual cell uncertainties in quadrature.

Figure \ref{fig:mass_evolution} compares the mass estimates obtained with the three methods, $M_{\mathrm{LTE}}$ (orange line), $M_{\mathrm{vir}}$ (blue line), and $M_{\mathrm{XF}}$ (green line), against the mass measured directly from the simulation, $M_{\mathrm{3D}}$. As shown, $M_{\mathrm{XF}}$ provides the closest agreement with the simulation-based mass, whereas $M_{\mathrm{vir}}$ systematically overestimates it by nearly one order of magnitude and $M_{\mathrm{LTE}}$ significantly underestimates it. We therefore adopt $M_{\mathrm{XF}}$ as our fiducial estimate of the pillar mass, and use it to derive the corresponding mean density of the structure (see Section~\ref{subsec:mass_density}).

\section{Second-order moment maps}
\label{sec:appendix_secondOrderMomentmaps}

\begin{figure*}
\includegraphics[width=0.43\textwidth]{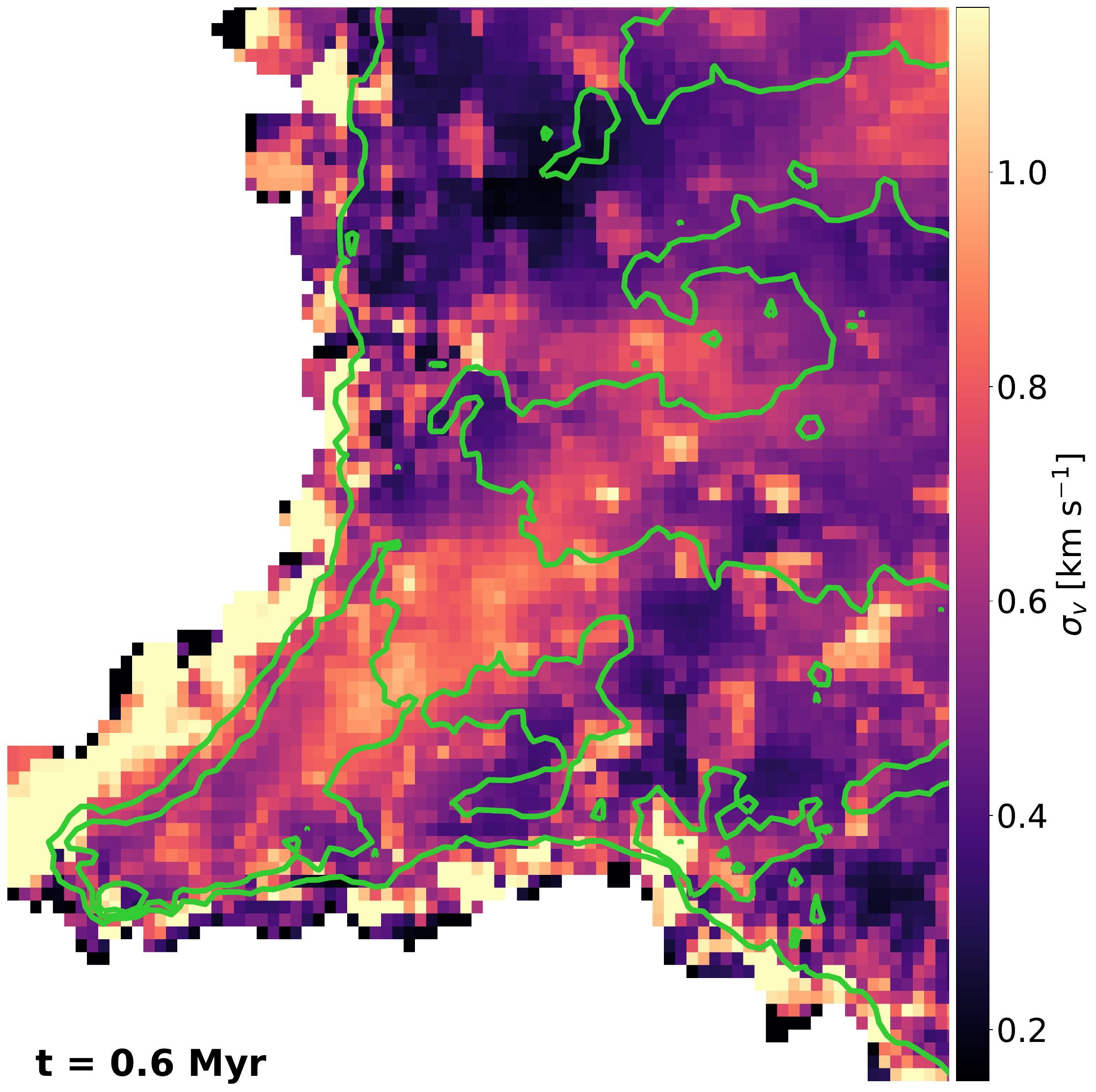}
\includegraphics[width=0.43\textwidth]{
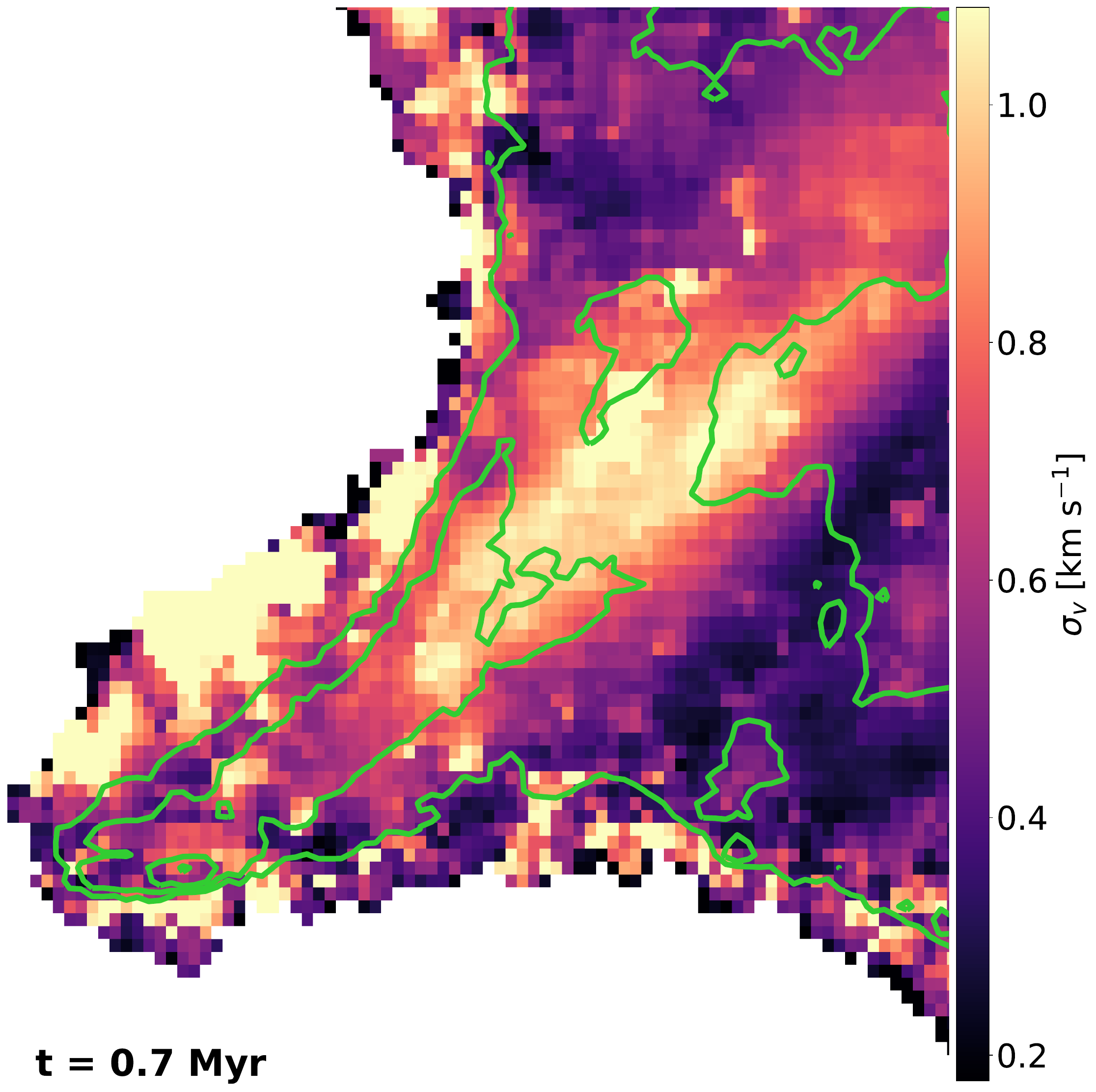}
\includegraphics[width=0.43\textwidth]{
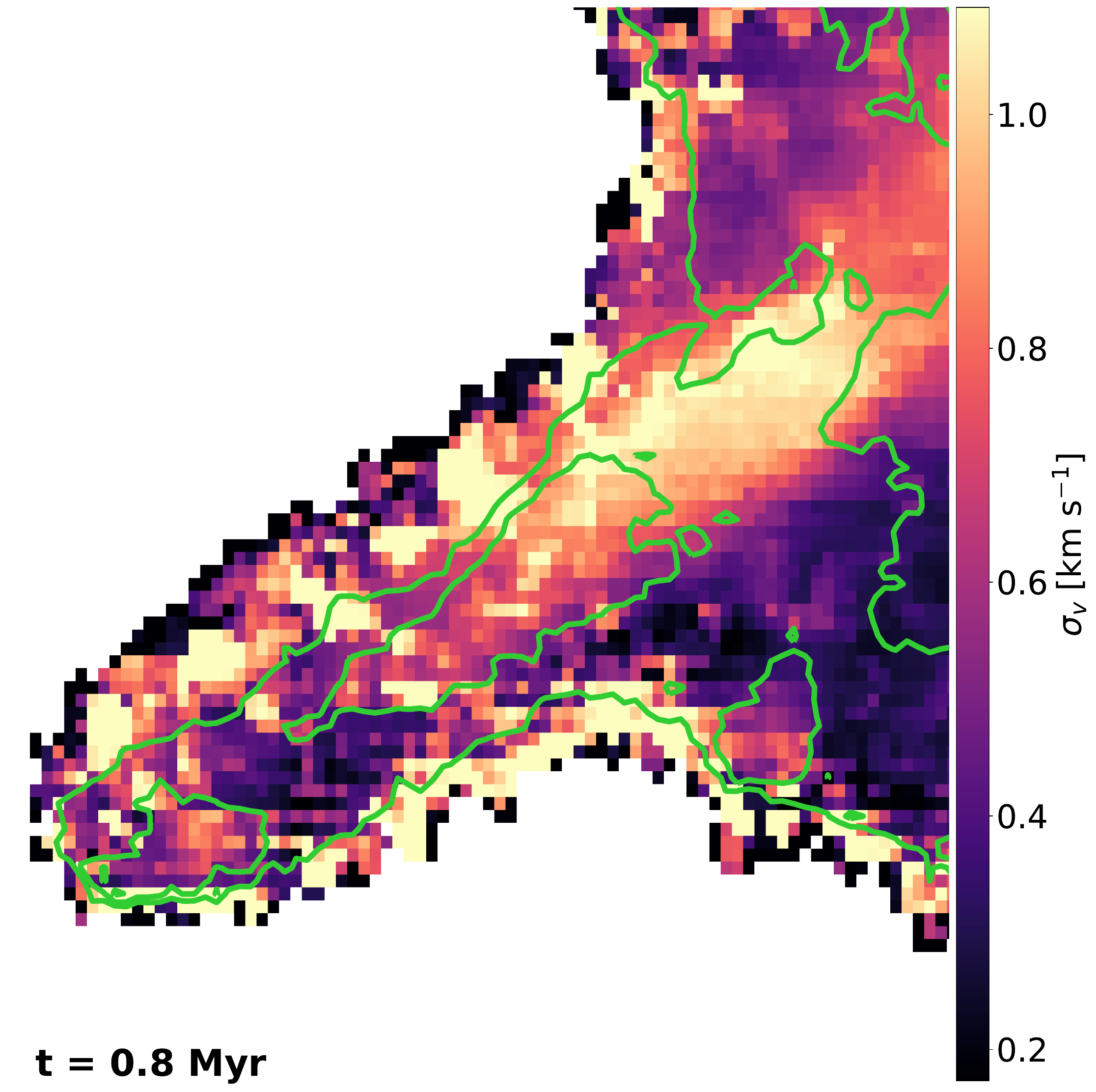}
\includegraphics[width=0.43\textwidth]{
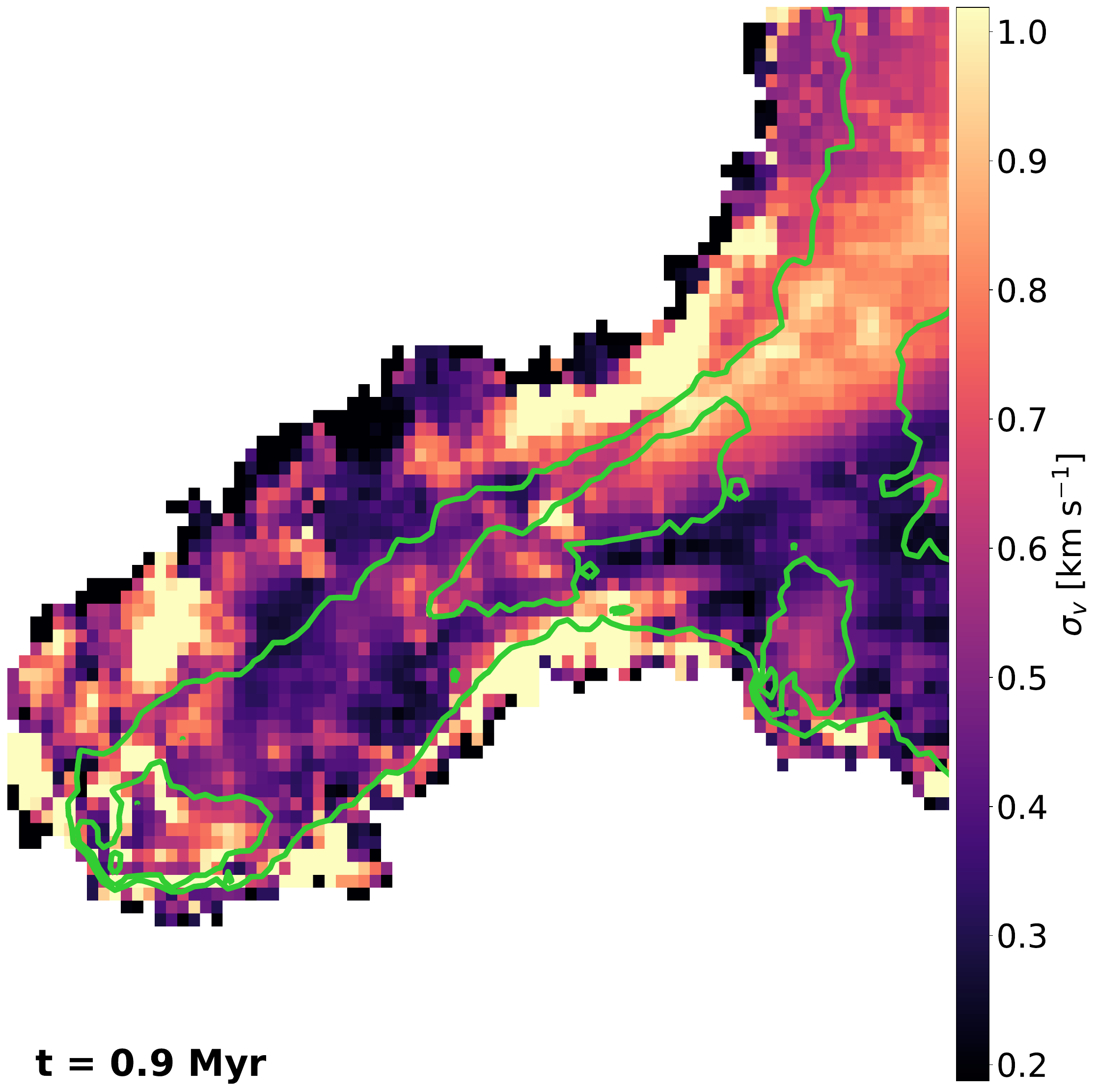}
\caption{ Evolution of the second-order moment maps of the pillar. The color-scale maps show the second spectral moment (velocity dispersion, $\sigma_v$) of the synthetic $^{13}\mathrm{CO}(J=1-0)$ emission during the stable phase of the pillar, from $t = 0.6$ to $0.9\,\mathrm{Myr}$. The overlaid green contours represent the zeroth moment (integrated intensity) of the same molecular transition at levels of $5$, $10$, $20$, and $40\,\mathrm{K\,km\,s^{-1}}$, outlining the dense gas structure. Compression-driven motions dominate the pillar, with the highest $\sigma_v$ concentrated along the dense spine and near the head.} 
\label{fig:secondordermoments}
\end{figure*}

To characterize the spatial distribution of the non-thermal gas motions and their relation to the dense gas morphology, we derived second-order spectral moment maps from the synthetic $^{13}\mathrm{CO}(J=1-0)$ PPV cubes. Figure~\ref{fig:secondordermoments} displays the time evolution of the velocity dispersion, $\sigma_v$, during the stable lifetime of the simulated pillar, from $t = 0.6$ to $0.9\,\mathrm{Myr}$. The overlaid green contours represent the integrated line intensity, or zeroth-order moment, and trace the dense morphological features of the structure.

These maps are used to derive the local line width, $\Delta v = \sqrt{8\ln 2}\,\sigma_v$, and the corresponding velocity dispersion used in the DCF-based estimates described in Section~\ref{subsec:velocity_dispersion}.

\section{Polarization angle distributions}
\label{sec:appendix_PolAngDist_gfits}
\begin{figure*}
    \centering
    \includegraphics[width=0.75\textwidth]{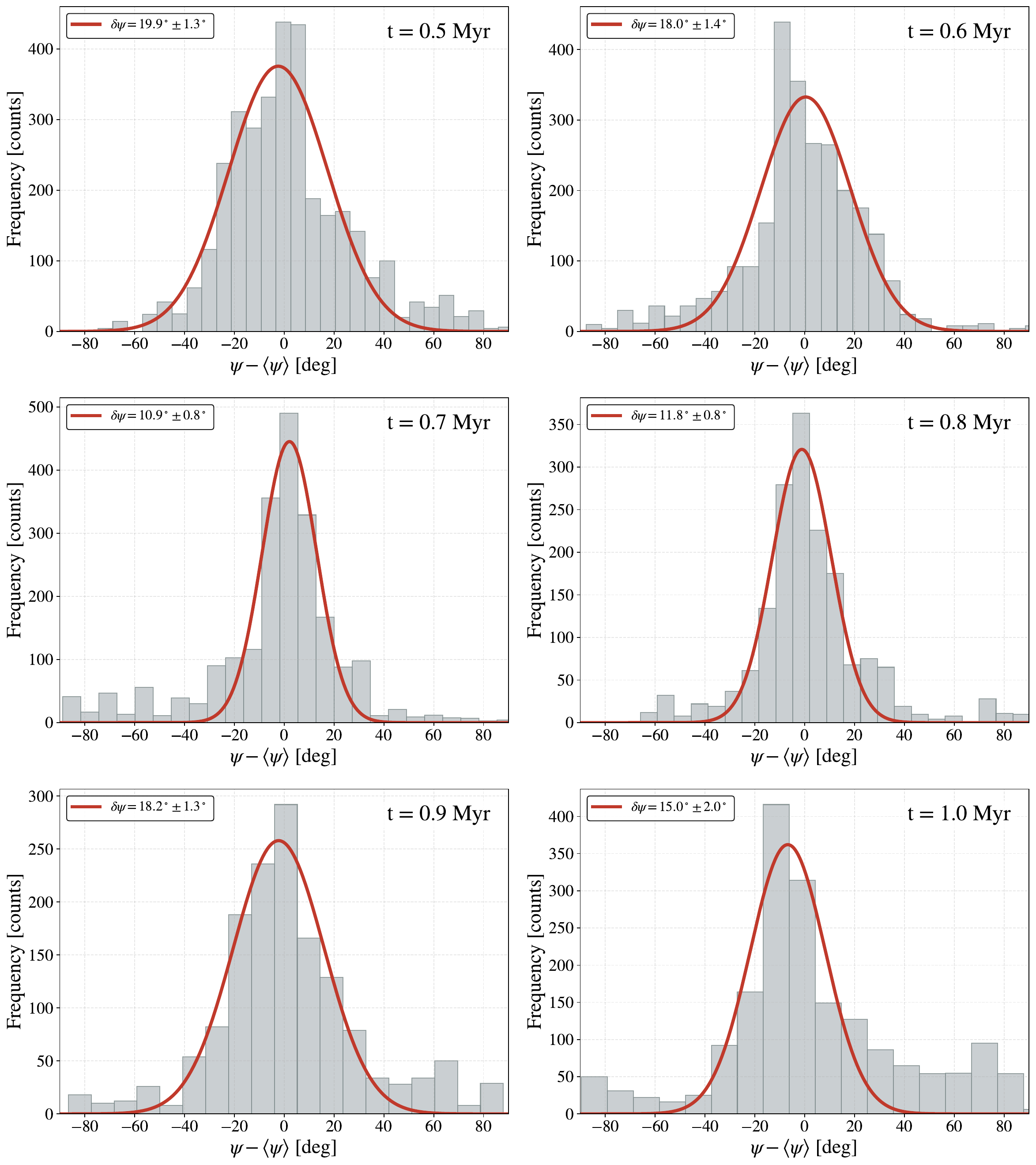} 
    \caption{Polarization-angle distributions and their Gaussian fits for the synthetic \texttt{POLARIS} observations. The distributions shown correspond to the full projected pillar without applying the polarimetric S/N masks, which is the fiducial configuration adopted in this work. The solid red line shows the best-fitting Gaussian model used to estimate the characteristic angular dispersion, $\delta\psi_{\mathrm{gfit}}$. Although the fitting procedure allows for two Gaussian components when statistically justified, nearly all timesteps are well described by a single Gaussian component; therefore, the single-Gaussian width is adopted as the main estimator of the angular dispersion. During the stable pillar phase, the distributions remain within the small-angle regime required for the DCF analysis}
    \label{fig:PAdist_gfits}
\end{figure*}

Figure~\ref{fig:PAdist_gfits} shows the polarization-angle distributions and their corresponding Gaussian fits at different timesteps. In contrast to a simple standard deviation of the polarization angles, which can be sensitive to extended tails or coherent large-scale components, we use the Gaussian width as the characteristic angular dispersion entering the DCF-based estimates. The fitting procedure follows the same motivation as the approach discussed by \citet{Polychronakis2025}, but in our case the distributions are predominantly unimodal, so a single Gaussian provides the appropriate description for nearly all snapshots.

The S/N-selected full-pillar and body-only S/N-selected measurements yield angular dispersions that are not significantly different from the fiducial values during the stable phase.

\section{Angular Dispersion Function (ADF) analysis}
\label{sec:appendix_adf_analysis}

To compare the angular dispersion obtained from our Gaussian fitting approach with an independent technique, we implemented the Angular Dispersion Function (ADF) method \citep{Hildebrand2009}. The ADF provides a scale-dependent estimate of the polarization-angle dispersion by measuring how rapidly the polarization angle changes as a function of projected spatial separation. In this work, the ADF is used as a complementary diagnostic, while the single-Gaussian fitting method is adopted as the main estimator of $\delta\psi$ used in the DCF-based magnetic-field calculations.

For a given region, we computed the squared angular difference $\Delta \Phi^2(l)$ for all pairs of pixels separated by a distance $l$, using the double-angle formulation to account for the $180^\circ$ ambiguity of the polarization vectors:
\begin{equation}
    \Delta \Phi =
    \frac{1}{2}
    \operatorname{atan2}
    \left[
        \sin(2\Delta\psi),
        \cos(2\Delta\psi)
    \right].
\end{equation}

Following \citet{Hildebrand2009}, we fitted the small-scale behavior of the ADF with a simple linear model,
\begin{equation}
    \langle \Delta \Phi^2(l) \rangle = b^2 + m^2 l^2 ,
\end{equation}
where the intercept $b^2$ traces the small-scale angular contribution, while the $m^2l^2$ term accounts for smoothly varying large-scale magnetic-field structure.

We found that a uniform fitting range of $l \leq 5$ pixels provides the most stable and physically consistent estimates for all configurations and timesteps. We therefore adopt $l \leq 5$ pixels, corresponding to $\sim 0.08\,\mathrm{pc}$ in our projected maps, for every ADF fit. This fixed small-scale range avoids contamination from the coherent pillar-scale magnetic curvature while keeping the ADF measurement consistent across the full unmasked pillar and the S/N-selected pillar body.

Because our synthetic observations represent an idealized case without telescope beam convolution smoothing, the intercept of the fit, $b^2$, is directly related to the turbulent angular dispersion without needing beam-size corrections. Therefore, the intrinsic dispersion is simply extracted as:
\begin{equation}
    \delta\psi = \sqrt{\frac{b^2}{2}}.
\end{equation}

We applied this ADF methodology to two distinct quality conditions: (1) the entire pillar without S/N filters, and (2) the entire pillar retaining only high S/N polarimetric data ($p/\delta p > 3$ and $I/\delta I > 10$).  The resulting ADF dispersions provide an independent check on the angular-dispersion measurements, but the main DCF and ST21 estimates reported in this work are based on the single-Gaussian widths, since the polarization-angle distributions are predominantly unimodal in our analysis.

\section{Robustness of the synthetic polarization morphology}
\label{sec:appendix_polaris_convergence}

To test the dependence of our morphological conclusions on the dust alignment efficiency and the numerical ray-tracing resolution, we performed a parametric study using the \texttt{POLARIS} radiative transfer code \citep{Reissl2016}. The objective of these experiments is to show that the highly ordered magnetic field morphology derived from our synthetic dust polarization maps is an intrinsic physical feature of the R-MHD simulation and not a geometric artifact of the assumed dust properties or spatial subsampling. These experiments were evaluated using the reduced chi-squared test ($\chi^2_{\nu}$) and the root-mean-square error (RMSE) with respect to the projected 3D magnetic field weighed by density.

Our fiducial setup uses the Radiative Torque alignment theory (\texttt{ALIG\_RAT}) with a sub-pixeling level of 1 (up to 4 sub-rays per pixel) and a fraction of high-$J$ state grains of $f_{\mathrm{high-}J} = 0.25$. We designed three additional sets of tests based on these fiducial conditions:
\begin{itemize}
    \item { {Test 1:} Evaluate variations of $f_{\mathrm{high-}J}$ ($0.10$ and $0.50$).}
    \item { {Test 2:} Introduces the effect of thermal fluctuations within the dust grains (imperfect internal alignment, \texttt{ALIG\_INTERNAL}) alongside an increased sub-pixeling level of 2 (up to 16 sub-rays per pixel) for different values of $f_{\mathrm{high-}J}$.}
    \item { {Test 3:} Increases the sub-pixeling level to 3 (up to 64 sub-rays per pixel) to rule out spatial subsampling artifacts, testing the same $f_{\mathrm{high-}J}$ variations.}
\end{itemize}

The physical impact of these parameters on emergent polarization can be understood through the Rayleigh reduction factor $R$, which quantifies the overall grain alignment efficiency. According to RAT alignment theory \citep{Andersson2015, Reissl2016}, when thermal fluctuations are neglected (Fiducial, Test 1, and Test 3), the code assumes perfect alignment ($R=1$) for all eligible grains, bypassing the parameter $f_{\mathrm{high-}J}$. Consequently, the mean polarization fraction $\langle p \rangle$ remains constant at its theoretical maximum of $\approx 2.43\%$. However, when imperfect internal alignment is activated (Test 2), the alignment efficiency scales realistically down, and $\langle p \rangle$ becomes highly dependent on $f_{\mathrm{high-}J}$, dropping to $1.09\%$ for $f_{\mathrm{high-}J} = 0.10$ and reaching $1.69\%$ for $f_{\mathrm{high-}J} = 0.50$.

Despite these significant changes in the total polarization fraction, the orientation of the polarization pseudo-vectors ($\psi$) remains mathematically invariant. As described by \citet{Reissl2016}, the polarization angle in the plane of the sky is strictly defined by the Stokes parameters $Q$ and $U$ (see Eq.~\ref{eq:polarisation_angle}).

Because the polarized intensities $Q$ and $U$ scale proportionally by the exact same factor $R$, this dependence cancels out completely in the quotient $U/Q$. As a result, the geometric morphology of the mapped magnetic field is independent of the assumed dust alignment efficiency.

This mathematical invariance is empirically confirmed by our metrics, summarized in Table~\ref{tab:polaris_experiments}. Across all variations of $f_{\mathrm{high-}J}$ and internal alignment modules, the RMSE and $\chi^2_{\nu}$ between the synthetic polarization angles and the density-weighted projected 3D magnetic field remain completely invariant at $31.43^{\circ}$ and $\sim 4.55$, respectively. Furthermore, increasing the ray-tracing resolution (sub-pixeling up to level 3) shows that the highly ordered magnetic field morphology and the relatively low angular dispersion ($\delta\psi$) observed in our maps are not artifacts of numerical smoothing.

Although $f_{\mathrm{high-}J}$ and sub-pixeling do not alter the polarization angle $\psi$, the orientation of the polarization vectors can still be modified by other physical conditions and parameters. These include: wavelength regime ($\lambda$), line of sight (LOS), and alternative alignment mechanisms \citep[see ][and references therein]{Reissl2016}.  

Although certain physical parameters, such as the observing wavelength or the viewing angle, can fundamentally alter the orientation of the polarization vectors $\psi$, exploring these variations is beyond the scope of this study. Our synthetic observations are strictly constrained to a wavelength of $850\,\mu\mathrm{m}$ to ensure a direct and consistent comparison with the measurements of Pillar II in M16 by \citetalias{Pattle2018}. In this sub-millimeter regime, ambiguities are avoided, as the polarization is purely due to thermal dust re-emission, resulting in polarization vectors that are strictly orthogonal to the magnetic field \citep{Andersson2015, Reissl2016}. Furthermore, because the magnetic field in our simulated pillar is highly ordered and aligned along its major axis due to external compression, moderate variations in the viewing angle would only affect the total polarization fraction, but cannot artificially generate the large characteristic polarization-angle dispersion.

In conclusion, our parametric study verifies that the morphological analysis of our synthetic maps at $850\,\mu\mathrm{m}$ is robust, consistently tracing the projected magnetic field weighed by density regardless of uncertainties in grain alignment efficiency.

\begin{table}
    \centering
    \caption{Summary of the \texttt{POLARIS} radiative transfer experiments testing the robustness of the synthetic polarization maps. The root-mean-square error (RMSE) and the reduced chi-squared ($\chi^2_{\nu}$) between the synthetic polarization angles and the density-weighted projected 3D magnetic field remain invariant regardless of the dust alignment efficiency or the ray-tracing resolution. Varying $f_{\mathrm{high-}J}$ only scales the mean polarization fraction $\langle p \rangle$ when imperfect internal alignment (thermal fluctuations) is considered.}
    \label{tab:polaris_experiments}
    \setlength{\tabcolsep}{3pt} 
    \begin{tabular}{l c c c c c c}
        \hline
        Exp. & Align. module & Sub-pix & $f_{\mathrm{high-}J}$ & $\langle p \rangle [\%] $ & $\chi^2_{\nu}$ & RMSE \\
        \hline
        Fiducial & \texttt{ALIG\_RAT} & 1 & 0.25 & 2.4319 & 4.555 & $31.43^{\circ}$ \\
        Test 1a & \texttt{ALIG\_RAT} & 1 & 0.10 & 2.4319 & 4.554 & $31.43^{\circ}$ \\
        Test 1b & \texttt{ALIG\_RAT} & 1 & 0.50 & 2.4319 & 4.555 & $31.43^{\circ}$ \\
        \hline
        Test 2a & \texttt{RAT} + \texttt{INTERNAL} & 2 & 0.01 & 1.0928 & 4.555 & $31.43^{\circ}$ \\
        Test 2b & \texttt{RAT} + \texttt{INTERNAL} & 2 & 0.25 & 1.3155 & 4.555 & $31.43^{\circ}$ \\
        Test 2c & \texttt{RAT} + \texttt{INTERNAL} & 2 & 0.50 & 1.6871 & 4.555 & $31.43^{\circ}$ \\
        \hline
        Test 3a & \texttt{ALIG\_RAT} & 3 & 0.10 & 2.4319 & 4.555 & $31.43^{\circ}$ \\
        Test 3b & \texttt{ALIG\_RAT} & 3 & 0.25 & 2.4319 & 4.555 & $31.43^{\circ}$ \\
        Test 3c & \texttt{ALIG\_RAT} & 3 & 0.50 & 2.4319 & 4.555 & $31.43^{\circ}$ \\
        \hline
    \end{tabular}
\end{table}


\bsp	
\label{lastpage}
\end{document}

%% file: definitions.tex
\newcommand{\HII}{\ion{H}{ii}}

\def\apjl{ApJL}
\def\aap{A\& A}

\def\apjs{ApJS}

%% file: refs.bib
@article{Gritschneder+2009,
doi = {10.1088/0004-637X/694/1/L26},
url = {https://doi.org/10.1088/0004-637X/694/1/L26},
year = {2009},
month = {feb},
publisher = {The American Astronomical Society},
volume = {694},
number = {1},
pages = {L26},
author = {Gritschneder, Matthias and Naab, Thorsten and Walch, Stefanie and Burkert, Andreas and Heitsch, Fabian},
title = {DRIVING TURBULENCE AND TRIGGERING STAR FORMATION BY IONIZING RADIATION},
journal = {The Astrophysical Journal}
}

@ARTICLE{Tremblin+2012,
       author = {{Tremblin}, P. and {Audit}, E. and {Minier}, V. and {Schneider}, N.},
        title = "{3D simulations of pillar formation around HII regions: the importance of shock curvature}",
      journal = {\aap},
         year = 2012,
        month = feb,
       volume = {538},
          eid = {A31},
        pages = {A31},
          doi = {10.1051/0004-6361/201118031},
archivePrefix = {arXiv},
       eprint = {1111.1522},
 primaryClass = {astro-ph.SR},
       adsurl = {https://ui.adsabs.harvard.edu/abs/2012A&A...538A..31T}
}

@ARTICLE{Sugitani+2007,
       author = {{Sugitani}, Koji and {Watanabe}, Makoto and {Tamura}, Motohide and {Kandori}, Ryo and {Hough}, James H. and {Nishiyama}, Shogo and {Nakajima}, Yasushi and {Kusakabe}, Nobuhiko and {Hashimoto}, Jun and {Nagayama}, Takahiro and {Nagashima}, Chie and {Kato}, Daisuke and {Fukuda}, Naoya},
        title = "{Near-Infrared Polarimetry of the Eagle Nebula (M 16)}",
      journal = {\pasj},
         year = 2007,
        month = jun,
       volume = {59},
        pages = {507-517},
          doi = {10.1093/pasj/59.3.507},
archivePrefix = {arXiv},
       eprint = {astro-ph/0611950},
 primaryClass = {astro-ph},
       adsurl = {https://ui.adsabs.harvard.edu/abs/2007PASJ...59..507S}
}

@ARTICLE{Bertoldi+1989,
       author = {{Bertoldi}, Frank},
        title = "{The Photoevaporation of Interstellar Clouds. I. Radiation-driven Implosion}",
      journal = {\apj},
         year = 1989,
        month = nov,
       volume = {346},
        pages = {735},
          doi = {10.1086/168055},
       adsurl = {https://ui.adsabs.harvard.edu/abs/1989ApJ...346..735B}
}

@ARTICLE{Waagan+11,
   author = {{Waagan}, K. and {Federrath}, C. and {Klingenberg}, C.},
    title = "{A robust numerical scheme for highly compressible magnetohydrodynamics: Nonlinear stability, implementation and tests}",
  journal = {Journal of Computational Physics},
archivePrefix = "arXiv",
   eprint = {1101.3007},
 primaryClass = "astro-ph.IM",
     year = 2011,
    month = may,
   volume = 230,
    pages = {3331-3351},
      doi = {10.1016/j.jcp.2011.01.026},
   adsurl = {http://adsabs.harvard.edu/abs/2011JCoPh.230.3331W}
}

@ARTICLE{Rijkhorst06,
   author = {{Rijkhorst}, E.-J. and {Plewa}, T. and {Dubey}, A. and {Mellema}, G.
	},
    title = "{Hybrid characteristics: 3D radiative transfer for parallel adaptive mesh refinement hydrodynamics}",
  journal = {\aap},
   eprint = {astro-ph/0505213},
     year = 2006,
    month = jun,
   volume = 452,
    pages = {907-920},
      doi = {10.1051/0004-6361:20053401},
   adsurl = {http://adsabs.harvard.edu/abs/2006A%26A...452..907R}
}

@ARTICLE{Wunsch+18,
       author = {{W{\"u}nsch}, R. and {Walch}, S. and {Dinnbier}, F. and {Whitworth}, A.},
        title = "{Tree-based solvers for adaptive mesh refinement code FLASH - I: gravity and optical depths}",
      journal = {\mnras},
         year = 2018,
        month = apr,
       volume = {475},
       number = {3},
        pages = {3393-3418},
          doi = {10.1093/mnras/sty015},
archivePrefix = {arXiv},
       eprint = {1708.06142},
 primaryClass = {astro-ph.IM},
       adsurl = {https://ui.adsabs.harvard.edu/abs/2018MNRAS.475.3393W}
}

@ARTICLE{Federrath+10,
   author = {{Federrath}, C. and {Banerjee}, R. and {Clark}, P.~C. and {Klessen}, R.~S.
	},
    title = "{Modeling Collapse and Accretion in Turbulent Gas Clouds: Implementation and Comparison of Sink Particles in AMR and SPH}",
  journal = {\apj},
archivePrefix = "arXiv",
   eprint = {1001.4456},
 primaryClass = "astro-ph.SR",
     year = 2010,
    month = apr,
   volume = 713,
    pages = {269-290},
      doi = {10.1088/0004-637X/713/1/269},
   adsurl = {http://adsabs.harvard.edu/abs/2010ApJ...713..269F}
}

@ARTICLE{FLASH1,
   author = {{Fryxell}, B. and {Olson}, K. and {Ricker}, P. and {Timmes}, F.~X. and 
	{Zingale}, M. and {Lamb}, D.~Q. and {MacNeice}, P. and {Rosner}, R. and 
	{Truran}, J.~W. and {Tufo}, H.},
    title = "{FLASH: An Adaptive Mesh Hydrodynamics Code for Modeling Astrophysical Thermonuclear Flashes}",
  journal = {\apjs},
     year = 2000,
    month = nov,
   volume = 131,
    pages = {273-334},
      doi = {10.1086/317361},
   adsurl = {http://adsabs.harvard.edu/abs/2000ApJS..131..273F}
}

@ARTICLE{Truelove+97,
   author = {{Truelove}, J.~K. and {Klein}, R.~I. and {McKee}, C.~F. and 
	{Holliman}, II, J.~H. and {Howell}, L.~H. and {Greenough}, J.~A.
	},
    title = "{The Jeans Condition: A New Constraint on Spatial Resolution in Simulations of Isothermal Self-gravitational Hydrodynamics}",
  journal = {\apjl},
     year = 1997,
    month = nov,
   volume = 489,
    pages = {L179-L183},
      doi = {10.1086/310975},
   adsurl = {http://adsabs.harvard.edu/abs/1997ApJ...489L.179T}
}

@ARTICLE{Henney+09,
       author = {{Henney}, William J. and {Arthur}, S. Jane and {de Colle}, Fabio and
         {Mellema}, Garrelt},
        title = "{Radiation-magnetohydrodynamic simulations of the photoionization of magnetized globules}",
      journal = {\mnras},
         year = 2009,
        month = sep,
       volume = {398},
       number = {1},
        pages = {157-175},
          doi = {10.1111/j.1365-2966.2009.15153.x},
archivePrefix = {arXiv},
       eprint = {0810.1531},
 primaryClass = {astro-ph},
       adsurl = {https://ui.adsabs.harvard.edu/abs/2009MNRAS.398..157H}
}

@ARTICLE{Mackey+10,
       author = {{Mackey}, Jonathan and {Lim}, Andrew J.},
        title = "{Dynamical models for the formation of elephant trunks in HII regions}",
      journal = {\mnras},
         year = 2010,
        month = apr,
       volume = {403},
       number = {2},
        pages = {714-730},
          doi = {10.1111/j.1365-2966.2009.16181.x},
archivePrefix = {arXiv},
       eprint = {0912.1499},
 primaryClass = {astro-ph.GA},
       adsurl = {https://ui.adsabs.harvard.edu/abs/2010MNRAS.403..714M}
}

@ARTICLE{polaris,
       author = {{Reissl}, S. and {Wolf}, S. and {Brauer}, R.},
        title = "{Radiative transfer with POLARIS. I. Analysis of magnetic fields through synthetic dust continuum polarization measurements}",
      journal = {\aap},
         year = "2016",
        month = "Sep",
       volume = {593},
          eid = {A87},
        pages = {A87},
          doi = {10.1051/0004-6361/201424930},
archivePrefix = {arXiv},
       eprint = {1604.05305},
 primaryClass = {astro-ph.IM},
       adsurl = {https://ui.adsabs.harvard.edu/abs/2016A&A...593A..87R}
}

@ARTICLE{Seifried+19,
       author = {{Seifried}, D. and {Walch}, S. and {Reissl}, S. and
         {Ib{\'a}{\~n}ez-Mej{\'\i}a}, J.~C.},
        title = "{SILCC-Zoom: Polarization and depolarization in molecular clouds}",
      journal = {\mnras},
         year = "2019",
        month = "Jan",
       volume = {482},
       number = {2},
        pages = {2697-2716},
          doi = {10.1093/mnras/sty2831},
archivePrefix = {arXiv},
       eprint = {1804.10157},
 primaryClass = {astro-ph.GA},
       adsurl = {https://ui.adsabs.harvard.edu/abs/2019MNRAS.482.2697S}
}

@ARTICLE{Williams07,
       author = {{Williams}, R.~J.~R.},
        title = "{MHD Ionization Fronts}",
      journal = {Astrophysics and Space Science Proceedings},
         year = 2007,
        month = jan,
       volume = {1},
        pages = {129},
          doi = {10.1007/978-1-4020-5425-9_7},
       adsurl = {https://ui.adsabs.harvard.edu/abs/2007ASSP....1..129W}
}

@ARTICLE{ZA,
       author = {{Zamora-Avil{\'e}s}, Manuel and {V{\'a}zquez-Semadeni}, Enrique and
         {Gonz{\'a}lez}, Ricardo F. and {Franco}, Jos{\'e} and
         {Shore}, Steven N. and {Hartmann}, Lee W. and
         {Ballesteros-Paredes}, Javier and {Banerjee}, Robi and
         {K{\"o}rtgen}, Bastian},
        title = "{Structure and expansion law of H II regions in structured molecular clouds}",
      journal = {\mnras},
         year = "2019",
        month = "Aug",
       volume = {487},
       number = {2},
        pages = {2200-2214},
          doi = {10.1093/mnras/stz1235},
archivePrefix = {arXiv},
       eprint = {1903.01644},
 primaryClass = {astro-ph.GA},
       adsurl = {https://ui.adsabs.harvard.edu/abs/2019MNRAS.487.2200Z}
}

@ARTICLE{KI00,
   author = {{Koyama}, H. and {Inutsuka}, S. ~I.},
    title = "{Molecular Cloud Formation in Shock-compressed Layers}",
  journal = {\apj},
   eprint = {astro-ph/9912509},
     year = 2000,
    month = apr,
   volume = 532,
    pages = {980-993},
      doi = {10.1086/308594},
   adsurl = {http://adsabs.harvard.edu/abs/2000ApJ...532..980K}
}

@ARTICLE{KI02,
   author = {{Koyama}, H. and {Inutsuka}, S. ~I.},
    title = "{An Origin of Supersonic Motions in Interstellar Clouds}",
  journal = {\apjl},
   eprint = {astro-ph/0112420},
     year = 2002,
    month = jan,
   volume = 564,
    pages = {L97-L100},
      doi = {10.1086/338978},
   adsurl = {http://adsabs.harvard.edu/abs/2002ApJ...564L..97K}
}

@ARTICLE{Juarez+2017,
       author = {{Ju{\'a}rez}, Carmen and {Girart}, Josep M. and {Zamora-Avil{\'e}s}, Manuel and {Tang}, Ya-Wen and {Koch}, Patrick M. and {Liu}, Hauyu Baobab and {Palau}, Aina and {Ballesteros-Paredes}, Javier and {Zhang}, Qizhou and {Qiu}, Keping},
        title = "{Magnetized Converging Flows toward the Hot Core in the Intermediate/High-mass Star-forming Region NGC 6334 V}",
      journal = {\apj},
         year = 2017,
        month = jul,
       volume = {844},
       number = {1},
          eid = {44},
        pages = {44},
          doi = {10.3847/1538-4357/aa78a6},
archivePrefix = {arXiv},
       eprint = {1706.03534},
 primaryClass = {astro-ph.SR},
       adsurl = {https://ui.adsabs.harvard.edu/abs/2017ApJ...844...44J}
}

@article{Mackey_Lim2011,
    author = {Mackey, Jonathan and Lim, Andrew J.},
    title = "{Effects of magnetic fields on photoionized pillars and globules}",
    journal = {Monthly Notices of the Royal Astronomical Society},
    volume = {412},
    number = {3},
    pages = {2079-2094},
    year = {2011},
    month = {04},
    issn = {0035-8711},
    doi = {10.1111/j.1365-2966.2010.18043.x},
    url = {https://doi.org/10.1111/j.1365-2966.2010.18043.x},
    eprint = {https://academic.oup.com/mnras/article-pdf/412/3/2079/3619409/mnras0412-2079.pdf},
}

@article{Williams+2001,
    author = {Williams, R.J.R. and Ward-Thompson, D. and Whitworth, A.P.},
    title = "{Hydrodynamics of photoionized columns in the Eagle Nebula, M 16}",
    journal = {Monthly Notices of the Royal Astronomical Society},
    volume = {327},
    number = {3},
    pages = {788-798},
    year = {2001},
    month = {11},
    issn = {0035-8711},
    doi = {10.1046/j.1365-8711.2001.04757.x},
    url = {https://doi.org/10.1046/j.1365-8711.2001.04757.x},
    eprint = {https://academic.oup.com/mnras/article-pdf/327/3/788/2876628/327-3-788.pdf},
}

@ARTICLE{Vazquez_Semadeni+19,
       author = {{V{\'a}zquez-Semadeni}, Enrique and {Palau}, Aina and {Ballesteros-Paredes}, Javier and {G{\'o}mez}, Gilberto C. and {Zamora-Avil{\'e}s}, Manuel},
        title = "{Global hierarchical collapse in molecular clouds. Towards a comprehensive scenario}",
      journal = {\mnras},
         year = 2019,
        month = dec,
       volume = {490},
       number = {3},
        pages = {3061-3097},
          doi = {10.1093/mnras/stz2736},
       adsurl = {https://ui.adsabs.harvard.edu/abs/2019MNRAS.490.3061V}
}

@ARTICLE{Field_65,
   author = {{Field}, G.~B.},
    title = "{Thermal Instability.}",
  journal = {\apj},
     year = 1965,
    month = aug,
   volume = 142,
    pages = {531},
      doi = {10.1086/148317},
   adsurl = {http://adsabs.harvard.edu/abs/1965ApJ...142..531F}
}

@article{ZamoraAvils2019,
  title = {Structure and expansion law of Hii regions in structured molecular clouds},
  volume = {487},
  ISSN = {1365-2966},
  url = {http://dx.doi.org/10.1093/mnras/stz1235},
  DOI = {10.1093/mnras/stz1235},
  number = {2},
  journal = {Monthly Notices of the Royal Astronomical Society},
  publisher = {Oxford University Press (OUP)},
  author = {Zamora-Avilés,  Manuel and Vázquez-Semadeni,  Enrique and González,  Ricardo F and Franco,  José and Shore,  Steven N and Hartmann,  Lee W and Ballesteros-Paredes,  Javier and Banerjee,  Robi and K\"{o}rtgen,  Bastian},
  year = {2019},
  month = may,
  pages = {2200–2214}
}

@article{Reissl2016,
  title = {Radiative transfer with POLARIS: I. Analysis of magnetic fields through synthetic dust continuum polarization measurements},
  volume = {593},
  ISSN = {1432-0746},
  url = {http://dx.doi.org/10.1051/0004-6361/201424930},
  DOI = {10.1051/0004-6361/201424930},
  journal = {Astronomy &amp; Astrophysics},
  publisher = {EDP Sciences},
  author = {Reissl,  S. and Wolf,  S. and Brauer,  R.},
  year = {2016},
  month = sep,
  pages = {A87}
}

@misc{pattle2022magnetic,
      title={Magnetic fields in star formation: from clouds to cores}, 
      author={Kate Pattle and Laura Fissel and Mehrnoosh Tahani and Tie Liu and Evangelia Ntormousi},
      year={2022},
      eprint={2203.11179},
      archivePrefix={arXiv},
      primaryClass={astro-ph.GA}
}

@article{Pattle2018,
  title = {First Observations of the Magnetic Field inside the Pillars of Creation: Results from the BISTRO Survey},
  volume = {860},
  ISSN = {2041-8213},
  url = {http://dx.doi.org/10.3847/2041-8213/aac771},
  DOI = {10.3847/2041-8213/aac771},
  number = {1},
  journal = {The Astrophysical Journal Letters},
  publisher = {American Astronomical Society},
  author = {Pattle,  Kate and Ward-Thompson,  Derek and Hasegawa,  Tetsuo and Bastien,  Pierre and Kwon,  Woojin and Lai,  Shih-Ping and Qiu,  Keping and Furuya,  Ray and Berry,  David},
  year = {2018},
  month = jun,
  pages = {L6}
}

@article{White1999,
       author = {{White}, G.~J. and {Nelson}, R.~P. and {Holland}, W.~S. and {Robson}, E.~I. and {Greaves}, J.~S. and {McCaughrean}, M.~J. and {Pilbratt}, G.~L. and {Balser}, D.~S. and {Oka}, T. and {Sakamoto}, S. and {Hasegawa}, T. and {McCutcheon}, W.~H. and {Matthews}, H.~E. and {Fridlund}, C.~V.~M. and {Tothill}, N.~F.~H. and {Huldtgren}, M. and {Deane}, J.~R.},
        title = "{The Eagle Nebula's fingers - pointers to the earliest stages of star formation?}",
      journal = {\aap},
         year = 1999,
        month = feb,
       volume = {342},
        pages = {233-256},
       adsurl = {https://ui.adsabs.harvard.edu/abs/1999A&A...342..233W}
}

@ARTICLE{Mathis_1983,
       author = {{Mathis}, J.~S. and {Mezger}, P.~G. and {Panagia}, N.},
        title = "{Interstellar radiation field and dust temperatures in the diffuse interstellar medium and in giant molecular clouds}",
      journal = {\aap},
         year = 1983,
        month = nov,
       volume = {128},
        pages = {212-229},
       adsurl = {https://ui.adsabs.harvard.edu/abs/1983A&A...128..212M}
}

@ARTICLE{Camps_2015,
       author = {{Camps}, Peter and {Misselt}, Karl and {Bianchi}, Simone and {Lunttila}, Tuomas and {Pinte}, Christophe and {Natale}, Giovanni and {Juvela}, Mika and {Fischera}, Joerg and {Fitzgerald}, Michael P. and {Gordon}, Karl and {Baes}, Maarten and {Steinacker}, J{\"u}rgen},
        title = "{Benchmarking the calculation of stochastic heating and emissivity of dust grains in the context of radiative transfer simulations}",
      journal = {\aap},
         year = 2015,
        month = aug,
       volume = {580},
          eid = {A87},
        pages = {A87},
          doi = {10.1051/0004-6361/201525998},
archivePrefix = {arXiv},
       eprint = {1506.05304},
 primaryClass = {astro-ph.IM},
       adsurl = {https://ui.adsabs.harvard.edu/abs/2015A&A...580A..87C}
}

@article{Ostriker2001,
  title = {Density,  Velocity,  and Magnetic Field Structure in Turbulent Molecular Cloud Models},
  volume = {546},
  ISSN = {1538-4357},
  url = {http://dx.doi.org/10.1086/318290},
  DOI = {10.1086/318290},
  number = {2},
  journal = {The Astrophysical Journal},
  publisher = {American Astronomical Society},
  author = {Ostriker,  Eve C. and Stone,  James M. and Gammie,  Charles F.},
  year = {2001},
  month = jan,
  pages = {980–1005}
}

@ARTICLE{McLeod2015,
       author = {{McLeod}, A.~F. and {Dale}, J.~E. and {Ginsburg}, A. and {Ercolano}, B. and {Gritschneder}, M. and {Ramsay}, S. and {Testi}, L.},
        title = "{The Pillars of Creation revisited with MUSE: gas kinematics and high-mass stellar feedback traced by optical spectroscopy}",
      journal = {\mnras},
         year = 2015,
        month = jun,
       volume = {450},
       number = {1},
        pages = {1057-1076},
          doi = {10.1093/mnras/stv680},
archivePrefix = {arXiv},
       eprint = {1504.03323},
 primaryClass = {astro-ph.SR},
       adsurl = {https://ui.adsabs.harvard.edu/abs/2015MNRAS.450.1057M}
}

@ARTICLE{Hildebrand2009,
       author = {{Hildebrand}, Roger H. and {Kirby}, Larry and {Dotson}, Jessie L. and {Houde}, Martin and {Vaillancourt}, John E.},
        title = "{Dispersion of Magnetic Fields in Molecular Clouds. I}",
      journal = {\apj},
         year = 2009,
        month = may,
       volume = {696},
       number = {1},
        pages = {567-573},
          doi = {10.1088/0004-637X/696/1/567},
archivePrefix = {arXiv},
       eprint = {0811.0813},
 primaryClass = {astro-ph},
       adsurl = {https://ui.adsabs.harvard.edu/abs/2009ApJ...696..567H}
}

@ARTICLE{Chen2022,
       author = {{Chen}, Che-Yu and {Li}, Zhi-Yun and {Mazzei}, Renato R. and {Park}, Jinsoo and {Fissel}, Laura M. and {Chen}, Michael C. -Y. and {Klein}, Richard I. and {Li}, Pak Shing},
        title = "{The Davis-Chandrasekhar-Fermi method revisited}",
      journal = {\mnras},
         year = 2022,
        month = aug,
       volume = {514},
       number = {2},
        pages = {1575-1594},
          doi = {10.1093/mnras/stac1417},
archivePrefix = {arXiv},
       eprint = {2205.09134},
 primaryClass = {astro-ph.GA},
       adsurl = {https://ui.adsabs.harvard.edu/abs/2022MNRAS.514.1575C}
}

@article{Skalidis2021sqrtofdPA,
   title={Why take the square root? An assessment of interstellar magnetic field strength estimation methods},
   volume={656},
   ISSN={1432-0746},
   url={http://dx.doi.org/10.1051/0004-6361/202142045},
   DOI={10.1051/0004-6361/202142045},
   journal={Astronomy &amp; Astrophysics},
   publisher={EDP Sciences},
   author={Skalidis, R. and Sternberg, J. and Beattie, J. R. and Pavlidou, V. and Tassis, K.},
   year={2021},
   month=dec, pages={A118} }

@article{Skalidis2021,
  title = {High-accuracy estimation of magnetic field strength in the interstellar medium from dust polarization},
  volume = {647},
  ISSN = {1432-0746},
  url = {http://dx.doi.org/10.1051/0004-6361/202039779},
  DOI = {10.1051/0004-6361/202039779},
  journal = {Astronomy &amp; Astrophysics},
  publisher = {EDP Sciences},
  author = {Skalidis,  Raphael and Tassis,  Konstantinos},
  year = {2021},
  month = mar,
  pages = {A186}
}

@article{Peters_2010,
   title={H II REGIONS: WITNESSES TO MASSIVE STAR FORMATION},
   volume={711},
   ISSN={1538-4357},
   url={http://dx.doi.org/10.1088/0004-637X/711/2/1017},
   DOI={10.1088/0004-637x/711/2/1017},
   number={2},
   journal={The Astrophysical Journal},
   publisher={American Astronomical Society},
   author={Peters, Thomas and Banerjee, Robi and Klessen, Ralf S. and Low, Mordecai-Mark Mac and Galván-Madrid, Roberto and Keto, Eric R.},
   year={2010},
   month=feb, pages={1017–1028} }

@article{Hoang2009,
  title = {GRAIN ALIGNMENT INDUCED BY RADIATIVE TORQUES: EFFECTS OF INTERNAL RELAXATION OF ENERGY AND COMPLEX RADIATION FIELD},
  volume = {697},
  ISSN = {1538-4357},
  url = {http://dx.doi.org/10.1088/0004-637X/697/2/1316},
  DOI = {10.1088/0004-637x/697/2/1316},
  number = {2},
  journal = {The Astrophysical Journal},
  publisher = {American Astronomical Society},
  author = {Hoang,  Thiem and Lazarian,  A.},
  year = {2009},
  month = may,
  pages = {1316–1333}
}

@article{chandrasekhar1953,
  title={Magnetic fields in spiral arms},
  author={Chandrasekhar, Subrahmanyan and Fermi, Enrico},
  journal={Astrophysical Journal},
  volume={118},
  pages={113--115},
  year={1953},
  publisher={American Astronomical Society}
}

@article{davis1951,
  title={The strength of interstellar magnetic fields},
  author={Davis Jr, Leverett},
  journal={Physical Review},
  volume={81},
  number={5},
  pages={890},
  year={1951},
  publisher={APS}
}

@article{hwang2023,
  title={Magnetic fields in the Horsehead Nebula},
  author={Hwang, Jihye and Pattle, Kate and Parsons, Harriet and Go, Mallory and Kim, Jongsoo},
  journal={The Astronomical Journal},
  volume={165},
  number={5},
  pages={198},
  year={2023},
  publisher={IOP Publishing}
}

@article{Walch2012,
  title={Dispersal of molecular clouds by ionizing radiation},
  author={Walch, SK and Whitworth, AP and Bisbas, T and W{\"u}nsch, R and Hubber, D},
  journal={Monthly Notices of the Royal Astronomical Society},
  volume={427},
  number={1},
  pages={625--636},
  year={2012},
  publisher={The Royal Astronomical Society}
}

@ARTICLE{Arthur2011,
       author = {{Arthur}, S.~J. and {Henney}, W.~J. and {Mellema}, G. and {de Colle}, F. and {V{\'a}zquez-Semadeni}, E.},
        title = "{Radiation-magnetohydrodynamic simulations of H II regions and their associated PDRs in turbulent molecular clouds}",
      journal = {\mnras},
         year = 2011,
        month = jun,
       volume = {414},
       number = {2},
        pages = {1747-1768},
          doi = {10.1111/j.1365-2966.2011.18507.x},
archivePrefix = {arXiv},
       eprint = {1101.5510},
 primaryClass = {astro-ph.GA},
       adsurl = {https://ui.adsabs.harvard.edu/abs/2011MNRAS.414.1747A}
}

@ARTICLE{Hester1996,
       author = {{Hester}, J.~J. and {Scowen}, P.~A. and {Sankrit}, R. and {Lauer}, T.~R. and {Ajhar}, E.~A. and {Baum}, W.~A. and {Code}, A. and {Currie}, D.~G. and {Danielson}, G.~E. and {Ewald}, S.~P. and {Faber}, S.~M. and {Grillmair}, C.~J. and {Groth}, E.~J. and {Holtzman}, J.~A. and {Hunter}, D.~A. and {Kristian}, J. and {Light}, R.~M. and {Lynds}, C.~R. and {Monet}, D.~G. and {O'Neil}, E.~J., Jr. and {Shaya}, E.~J. and {Seidelmann}, P.~K. and {Westphal}, J.~A.},
        title = "{Hubble Space Telescope WFPC2 Imaging of M16: Photoevaporation and Emerging Young Stellar Objects}",
      journal = {\aj},
         year = 1996,
        month = jun,
       volume = {111},
        pages = {2349},
          doi = {10.1086/117968},
       adsurl = {https://ui.adsabs.harvard.edu/abs/1996AJ....111.2349H}
}

@article{mathis1977,
  title={The size distribution of interstellar grains},
  author={Mathis, John S and Rumpl, William and Nordsieck, Kenneth H},
  journal={Astrophysical Journal, Part 1, vol. 217, Oct. 15, 1977, p. 425-433. NSF-supported research.},
  volume={217},
  pages={425--433},
  year={1977}
}

@article{draine2001,
  title={Infrared emission from interstellar dust. I. Stochastic heating of small grains},
  author={Draine, BT and Li, Aigen},
  journal={The Astrophysical Journal},
  volume={551},
  number={2},
  pages={807},
  year={2001},
  publisher={IOP Publishing}
}

@ARTICLE{LQZ22,
       author = {{Liu}, Junhao and {Qiu}, Keping and {Zhang}, Qizhou},
        title = "{Magnetic Fields in Star Formation: A Complete Compilation of All the DCF Estimations}",
      journal = {\apj},
         year = 2022,
        month = jan,
       volume = {925},
       number = {1},
          eid = {30},
        pages = {30},
          doi = {10.3847/1538-4357/ac3911},
archivePrefix = {arXiv},
       eprint = {2111.05836},
 primaryClass = {astro-ph.GA},
       adsurl = {https://ui.adsabs.harvard.edu/abs/2022ApJ...925...30L}
}

@article{liu2021calibrating,
  title={Calibrating the Davis--Chandrasekhar--Fermi Method with Numerical Simulations: Uncertainties in Estimating the Magnetic Field Strength from Statistics of Field Orientations},
  author={Liu, Junhao and Zhang, Qizhou and Commercon, Benoit and Valdivia, Valeska and Maury, Ana{\"e}lle and Qiu, Keping},
  journal={The Astrophysical Journal},
  volume={919},
  number={2},
  pages={79},
  year={2021},
  publisher={IOP Publishing}
}

@ARTICLE{GarciaSegura1996ApJ,
       author = {{Garcia-Segura}, Guillermo and {Franco}, Jose},
        title = "{From Ultracompact to Extended H II Regions}",
      journal = {\apj},
         year = 1996,
        month = sep,
       volume = {469},
        pages = {171},
          doi = {10.1086/177769},
       adsurl = {https://ui.adsabs.harvard.edu/abs/1996ApJ...469..171G}
}

@ARTICLE{Gritschneder2010,
       author = {{Gritschneder}, Matthias and {Burkert}, Andreas and {Naab}, Thorsten and {Walch}, Stefanie},
        title = "{Detailed Numerical Simulations on the Formation of Pillars Around H II Regions}",
      journal = {\apj},
         year = 2010,
        month = nov,
       volume = {723},
       number = {2},
        pages = {971-984},
          doi = {10.1088/0004-637X/723/2/971},
archivePrefix = {arXiv},
       eprint = {1009.0011},
 primaryClass = {astro-ph.SR},
       adsurl = {https://ui.adsabs.harvard.edu/abs/2010ApJ...723..971G}
}

@ARTICLE{Krumholz2007ApJ,
       author = {{Krumholz}, Mark R. and {Stone}, James M. and {Gardiner}, Thomas A.},
        title = "{Magnetohydrodynamic Evolution of H II Regions in Molecular Clouds: Simulation Methodology, Tests, and Uniform Media}",
      journal = {\apj},
         year = 2007,
        month = dec,
       volume = {671},
       number = {1},
        pages = {518-535},
          doi = {10.1086/522665},
archivePrefix = {arXiv},
       eprint = {astro-ph/0606539},
 primaryClass = {astro-ph},
       adsurl = {https://ui.adsabs.harvard.edu/abs/2007ApJ...671..518K}
}

@article{jackson2003catalog,
  title={A Catalog of Bright Filamentary Structures in the InterstellarMedium},
  author={Jackson, Tom and Werner, Michael and Gautier III, TN},
  journal={The Astrophysical Journal Supplement Series},
  volume={149},
  number={2},
  pages={365},
  year={2003},
  publisher={IOP Publishing}
}

@article{giuliani1979hydrodynamic,
  title={The hydrodynamic stability of ionization-shock fronts-Linear theory},
  author={Giuliani Jr, John L},
  journal={Astrophysical Journal, Part 1, vol. 233, Oct. 1, 1979, p. 280-293.},
  volume={233},
  pages={280--293},
  year={1979}
}

@article{vandervoort1962stability,
  title={On the Stability of Ionization Fronts.},
  author={Vandervoort, Peter O},
  journal={Astrophysical Journal, vol. 135, p. 212},
  volume={135},
  pages={212},
  year={1962}
}

@article{williams2002instability,
  title={On the instability of D-type ionization fronts},
  author={Williams, RJR},
  journal={Monthly Notices of the Royal Astronomical Society},
  volume={331},
  number={3},
  pages={693--706},
  year={2002},
  publisher={Blackwell Science, Ltd Oxford, UK}
}

@article{kim2014instability,
  title={Instability of Magnetized Ionization Fronts Surrounding H II Regions},
  author={Kim, Jeong-Gyu and Kim, Woong-Tae},
  journal={The Astrophysical Journal},
  volume={797},
  number={2},
  pages={135},
  year={2014},
  publisher={IOP Publishing}
}

@article{whalen2006multistep,
  title={A multistep algorithm for the radiation hydrodynamical transport of cosmological ionization fronts and ionized flows},
  author={Whalen, Daniel and Norman, Michael L},
  journal={The Astrophysical Journal Supplement Series},
  volume={162},
  number={2},
  pages={281},
  year={2006},
  publisher={IOP Publishing}
}

@article{garcia1996dynamical,
  title={The dynamical evolution of circumstellar gas around massive stars. I. The impact of the time sequence Ostar-> LBV-> WR star.},
  author={Garcia-Segura, G and Mac Low, M-M and Langer, N},
  journal={Astronomy and Astrophysics, v. 305, p. 229},
  volume={305},
  pages={229},
  year={1996}
}

@article{garcia1999shaping,
  title={Shaping bipolar and elliptical planetary nebulae: effects of stellar rotation, photoionization heating, and magnetic fields},
  author={Garc{\'\i}a-Segura, Guillermo and Langer, Norbert and Ro{\.z}yczka, Micha{\l} and Franco, Jose},
  journal={The Astrophysical Journal},
  volume={517},
  number={2},
  pages={767},
  year={1999},
  publisher={IOP Publishing}
}

@article{tenorio1991evolution,
  title={On the evolution of supernova remnants--II. Two-dimensional calculations of explosions inside pre-existing wind-driven bubbles},
  author={Tenorio-Tagle, G and R{\'o}{\.z}yczka, M and Franco, J and Bodenheimer, P},
  journal={Monthly Notices of the Royal Astronomical Society},
  volume={251},
  number={2},
  pages={318--329},
  year={1991},
  publisher={Oxford University Press Oxford, UK}
}

@article{franco1991evolution,
  title={Evolution of Supernova Remnant Inside Preexisting Wind-Driven Cavities},
  author={Franco, J and Tenorio-Tagle, G and Bodenheimer, P and Rozyczka, M},
  journal={Publications of the Astronomical Society of the Pacific},
  volume={103},
  number={666},
  pages={803},
  year={1991},
  publisher={IOP Publishing}
}

@article{karim2025sofiaM16,
  title={SOFIA FEEDBACK Survey: The Eagle Nebula in [C II] and Molecular Lines},
  author={Karim, Ramsey L and Pound, Marc W and Tielens, Alexander GGM and Kaastra, Jelle S and Townsley, Leisa K and Broos, Patrick S and Tiwari, Maitraiyee and Bonne, Lars and Kavak, {\"U}mit and Wolfire, Mark G and others},
  journal={arXiv preprint arXiv:2511.03978},
  year={2025}
}

@article{karim2023sofiaPillars,
  title={SOFIA FEEDBACK survey: the pillars of creation in [C II] and molecular lines},
  author={Karim, Ramsey L and Pound, Marc W and Tielens, Alexander GGM and Tiwari, Maitraiyee and Bonne, Lars and Wolfire, Mark G and Schneider, Nicola and Kavak, {\"U}mit and Mundy, Lee G and Simon, Robert and others},
  journal={The Astronomical Journal},
  volume={166},
  number={6},
  pages={240},
  year={2023},
  publisher={IOP Publishing}
}

@article{santillan1999collisions,
  title={The collisions of high-velocity clouds with a magnetized gaseous galactic disk},
  author={Santill{\'a}n, Alfredo and Franco, Jos{\'e} and Martos, Marco and Kim, Jongsoo},
  journal={The Astrophysical Journal},
  volume={515},
  number={2},
  pages={657},
  year={1999},
  publisher={IOP Publishing}
}

@article{hartquist1993clumps,
  title={Clumps their tails and the global sources they wag},
  author={Hartquist, TW and Dyson, JE},
  journal={Quarterly Journal of the Royal Astronomical Society, Vol. 34, NO. 1/MAR, P. 57, 1993},
  volume={34},
  pages={57},
  year={1993}
}

@article{remy2014dustgasmassratios,
  title={Gas-to-dust mass ratios in local galaxies over a 2 dex metallicity range},
  author={R{\'e}my-Ruyer, Aur{\'e}lie and Madden, SC and Galliano, F and Galametz, M and Takeuchi, TT and Asano, RS and Zhukovska, S and Lebouteiller, V and Cormier, D and Jones, A and others},
  journal={Astronomy \& Astrophysics},
  volume={563},
  pages={A31},
  year={2014},
  publisher={EDP Sciences}
}

@article{roman2022dust2gasratio_metal,
  title={METAL: the metal evolution, transport, and abundance in the large Magellanic Cloud Hubble Program. IV. Calibration of dust depletions versus abundance ratios in the Milky Way and Magellanic clouds and application to damped Ly$\alpha$ systems},
  author={Roman-Duval, Julia and Jenkins, Edward B and Tchernyshyov, Kirill and Clark, Christopher JR and De Cia, Annalisa and Gordon, Karl D and Hamanowicz, Aleksandra and Lebouteiller, Vianney and Rafelski, Marc and Sandstrom, Karin and others},
  journal={The Astrophysical Journal},
  volume={935},
  number={2},
  pages={105},
  year={2022},
  publisher={IOP Publishing}
}

@article{franco2025pressure_dust2gasmssratio,
  title={Pressure-regulated formation of molecular clouds and stars: the case of the Milky Way},
  author={Franco, Jos{\'e} and Rodr{\'\i}guez-Puebla, Aldo and Ballesteros-Paredes, Javier and Zamora-Avilez, Manuel},
  journal={Monthly Notices of the Royal Astronomical Society},
  volume={543},
  number={3},
  pages={2507--2522},
  year={2025},
  publisher={Oxford University Press}
}

@article{heitsch2001_DCFtestsInNumSims,
  title={Magnetic field diagnostics based on far-infrared polarimetry: tests using numerical simulations},
  author={Heitsch, Fabian and Zweibel, Ellen G and Mac Low, Mordecai-Mark and Li, Pakshing and Norman, Michael L},
  journal={The Astrophysical Journal},
  volume={561},
  number={2},
  pages={800},
  year={2001},
  publisher={IOP Publishing}
}

@article{Brinch2010,
  title={LIME--a flexible, non-LTE line excitation and radiation transfer method for millimeter and far-infrared wavelengths},
  author={Brinch, C and Hogerheijde, MR},
  journal={Astronomy \& Astrophysics},
  volume={523},
  pages={A25},
  year={2010},
  publisher={EDP Sciences}
}

@article{Sofue2020,
  title={CO line and radio continuum study of elephant trunks: the Pillars of Creation in M16},
  author={Sofue, Yoshiaki},
  journal={Monthly Notices of the Royal Astronomical Society},
  volume={492},
  number={4},
  pages={5966--5979},
  year={2020},
  publisher={Oxford University Press}
}

@article{polychronakis2025,
  title={A three-step approach to reliably estimate magnetic field strengths in star-forming regions},
  author={Polychronakis, Aristeidis and Tritsis, Aris and Skalidis, Raphael and Tassis, Konstantinos},
  journal={Astronomy \& Astrophysics},
  volume={700},
  pages={A256},
  year={2025},
  publisher={EDP Sciences}
}

@article{Andersson2015,
  title={Interstellar dust grain alignment},
  author={Andersson, BG and Lazarian, A and Vaillancourt, John E},
  journal={Annual Review of Astronomy and Astrophysics},
  volume={53},
  number={1},
  pages={501--539},
  year={2015},
  publisher={Annual Reviews}
}

@article{RetesRomero2017,
  title={The star-formation law in galactic high-mass star-forming molecular clouds},
  author={Retes-Romero, R and Mayya, YD and Luna, A and Carrasco, L},
  journal={The Astrophysical Journal},
  volume={839},
  number={2},
  pages={113},
  year={2017},
  publisher={The American Astronomical Society}
}

@article{kohno2024co,
  title={The CO-to-H2 conversion factor of Galactic giant molecular clouds using CO isotopologues: high-resolution X CO maps},
  author={Kohno, Mikito and Sofue, Yoshiaki},
  journal={Monthly Notices of the Royal Astronomical Society},
  volume={527},
  number={3},
  pages={9290--9302},
  year={2024},
  publisher={Oxford University Press}
}

@article{bolatto2013co,
  title={The CO-to-H2 conversion factor},
  author={Bolatto, Alberto D and Wolfire, Mark and Leroy, Adam K},
  journal={Annual Review of Astronomy and Astrophysics},
  volume={51},
  pages={207--268},
  year={2013},
  publisher={Annual Reviews}
}

@article{narayanan2012,
  title={A general model for the CO--H2 conversion factor in galaxies with applications to the star formation law},
  author={Narayanan, Desika and Krumholz, Mark R and Ostriker, Eve C and Hernquist, Lars},
  journal={Monthly Notices of the Royal Astronomical Society},
  volume={421},
  number={4},
  pages={3127--3146},
  year={2012},
  publisher={Blackwell Publishing Ltd Oxford, UK}
}

@article{Palau2021,
  title={Does the magnetic field suppress fragmentation in massive dense cores?},
  author={Palau, Aina and Zhang, Qizhou and Girart, Josep M and Liu, Junhao and Rao, Ramprasad and Koch, Patrick M and Estalella, Robert and Chen, Huei-Ru Vivien and Baobab Liu, Hauyu and Qiu, Keping and others},
  journal={The Astrophysical Journal},
  volume={912},
  number={2},
  pages={159},
  year={2021},
  publisher={The American Astronomical Society}
}

@ARTICLE{liu2019ApJ,
       author = {{Liu}, Junhao and {Qiu}, Keping and {Berry}, David and {Di Francesco}, James and {Bastien}, Pierre and {Koch}, Patrick M. and {Furuya}, Ray S. and {Kim}, Kee-Tae and {Coud{\'e}}, Simon and {Lee}, Chang Won and {Soam}, Archana and {Eswaraiah}, Chakali and {Li}, Di and {Hwang}, Jihye and {Lyo}, A.-Ran and {Pattle}, Kate and {Hasegawa}, Tetsuo and {Kwon}, Woojin and {Lai}, Shih-Ping and {Ward-Thompson}, Derek and {Ching}, Tao-Chung and {Chen}, Zhiwei and {Gu}, Qilao and {Li}, Dalei and {Li}, Hua-bai and {Liu}, Hong-Li and {Qian}, Lei and {Wang}, Hongchi and {Yuan}, Jinghua and {Zhang}, Chuan-Peng and {Zhang}, Guoyin and {Zhang}, Ya-Peng and {Zhou}, Jianjun and {Zhu}, Lei and {Andr{\'e}}, Philippe and {Arzoumanian}, Doris and {Aso}, Yusuke and {Byun}, Do-Young and {Chen}, Michael Chun-Yuan and {Chen}, Huei-Ru Vivien and {Chen}, Wen Ping and {Cho}, Jungyeon and {Choi}, Minho and {Chrysostomou}, Antonio and {Chung}, Eun Jung and {Doi}, Yasuo and {Drabek-Maunder}, Emily and {Dowell}, C. Darren and {Eyres}, Stewart P.~S. and {Falle}, Sam and {Fanciullo}, Lapo and {Fiege}, Jason and {Franzmann}, Erica and {Friberg}, Per and {Friesen}, Rachel K. and {Fuller}, Gary and {Gledhill}, Tim and {Graves}, Sarah F. and {Greaves}, Jane S. and {Griffin}, Matt J. and {Han}, Ilseung and {Hatchell}, Jennifer and {Hayashi}, Saeko S. and {Hoang}, Thiem and {Holland}, Wayne and {Houde}, Martin and {Inoue}, Tsuyoshi and {Inutsuka}, Shu-ichiro and {Iwasaki}, Kazunari and {Jeong}, Il-Gyo and {Johnstone}, Doug and {Kanamori}, Yoshihiro and {Kang}, Ji-hyun and {Kang}, Miju and {Kang}, Sung-ju and {Kataoka}, Akimasa and {Kawabata}, Koji S. and {Kemper}, Francisca and {Kim}, Gwanjeong and {Kim}, Jongsoo and {Kim}, Kyoung Hee and {Kim}, Mi-Ryang and {Kim}, Shinyoung and {Kirk}, Jason M. and {Kobayashi}, Masato I.~N. and {Kusune}, Takayoshi and {Kwon}, Jungmi and {Lacaille}, Kevin M. and {Lee}, Chin-Fei and {Lee}, Jeong-Eun and {Lee}, Hyeseung and {Lee}, Sang-Sung and {Liu}, Sheng-Yuan and {Liu}, Tie and {van Loo}, Sven and {Mairs}, Steve and {Matsumura}, Masafumi and {Matthews}, Brenda C. and {Moriarty-Schieven}, Gerald H. and {Nagata}, Tetsuya and {Nakamura}, Fumitaka and {Nakanishi}, Hiroyuki and {Ohashi}, Nagayoshi and {Onaka}, Takashi and {Parker}, Josh and {Parsons}, Harriet and {Pascale}, Enzo and {Peretto}, Nicolas and {Pon}, Andy and {Pyo}, Tae-Soo and {Rao}, Ramprasad and {Rawlings}, Mark G. and {Retter}, Brendan and {Richer}, John and {Rigby}, Andrew and {Robitaille}, Jean-Fran{\c{c}}ois and {Sadavoy}, Sarah and {Saito}, Hiro and {Savini}, Giorgio and {Scaife}, Anna M.~M. and {Seta}, Masumichi and {Shinnaga}, Hiroko and {Tamura}, Motohide and {Tang}, Ya-Wen and {Tomisaka}, Kohji and {Tsukamoto}, Yusuke and {Wang}, Jia-Wei and {Whitworth}, Anthony P. and {Yen}, Hsi-Wei and {Yoo}, Hyunju and {Zenko}, Tetsuya},
        title = "{The JCMT BISTRO Survey: The Magnetic Field in the Starless Core {\ensuremath{\rho}} Ophiuchus C}",
      journal = {\apj},
         year = 2019,
        month = may,
       volume = {877},
       number = {1},
          eid = {43},
        pages = {43},
          doi = {10.3847/1538-4357/ab0958},
archivePrefix = {arXiv},
       eprint = {1902.07734},
 primaryClass = {astro-ph.GA},
       adsurl = {https://ui.adsabs.harvard.edu/abs/2019ApJ...877...43L}
}
